\documentclass[sigconf,preprint]{acmart}

\usepackage{multirow}
\usepackage[breakable]{tcolorbox}
\usepackage{float}  
\usepackage{placeins}  

\AtBeginDocument{%
  }

\setcopyright{acmlicensed}
\copyrightyear{2026}
\acmYear{2026}
\acmDOI{XXXXXXX.XXXXXXX}

\acmConference[Preprint]{Preprint -- This manuscript is under review and has not yet been published at a venue}{2026}{Anywhere}
\acmISBN{978-1-4503-XXXX-X/2026/08}

\newcommand{\benchmark}{\textit{ISD-Agent-Bench}}
\newcommand{\numscenarios}{25,795}
\newcommand{\numtest}{1,202}

\graphicspath{{figures/}}

\begin{document}

\title{ISD-Agent-Bench: A Comprehensive Benchmark for Evaluating LLM-based Instructional Design Agents}

\author{YoungHoon Jeon$^{1}$, Suwan Kim$^{1}$, Haein Son$^{1}$, Sookbun Lee$^{2\dagger}$, Yeil Jeong$^{3\dagger}$, Unggi Lee$^{4\dagger}$}
\affiliation{%
  \institution{$^{1}$Upstage, $^{2}$Opentutorials, $^{3}$Indiana University Bloomington, $^{4}$Korea University Sejong Campus}
  \institution{\{yesica, suwankim, henny\}@upstage.ai, blackdew7@gmail.com \\ yeilj@iu.edu, codingchild@korea.ac.kr}
  \country{}}

\renewcommand{\shortauthors}{Jeon et al.}

\begin{abstract}
Large Language Model (LLM) agents have shown promising potential in automating Instructional Systems Design (ISD), a systematic approach to developing educational programs. However, evaluating these agents remains challenging due to the lack of standardized benchmarks and the risk of LLM-as-judge bias. We present ISD-Agent-Bench, a comprehensive benchmark comprising 25,795 scenarios generated via a Context Matrix framework that combines 51 contextual variables across 5 categories with 33 ISD sub-steps derived from the ADDIE model. To ensure evaluation reliability, we employ a multi-judge protocol using diverse LLMs from different providers, achieving high inter-judge reliability. We compare existing ISD agents with novel agents grounded in classical ISD theories such as ADDIE, Dick \& Carey, and Rapid Prototyping ISD. Experiments on 1,017 test scenarios demonstrate that integrating classical ISD frameworks with modern ReAct-style reasoning achieves the highest performance, outperforming both pure theory-based agents and technique-only approaches. Further analysis reveals that theoretical quality strongly correlates with benchmark performance, with theory-based agents showing significant advantages in problem-centered design and objective-assessment alignment. Our work provides a foundation for systematic LLM-based ISD research.
\end{abstract}



\maketitle

\section{Introduction}
\label{sec:introduction}

Instructional Systems Design (ISD) is a systematic process for creating effective learning experiences, encompassing five phases: Analysis, Design, Development, Implementation, and Evaluation, commonly known as the ADDIE model~\cite{branson1975addie}. ISD determines the quality and effectiveness of educational programs across diverse contexts, from corporate training to academic curricula. However, ISD is inherently a complex, multi-step task that requires human experts to execute each phase carefully. The process demands collaboration among instructional designers, subject matter experts, and multimedia specialists, with project timelines typically spanning weeks to months depending on scope and complexity. This significant investment of time and resources limits the scalability of high-quality ISD.

Recent advances in Large Language Models (LLMs) have sparked interest in applying them to ISD tasks. Hu et al.~\cite{hu2024teachingplan} explored GPT-4's potential for teaching plan generation and evaluation, while LessonPlanner~\cite{lessonplanner2024} assists novice teachers in preparing pedagogy-driven lesson plans. However, these approaches remain at the level of \textit{assisting} human designers rather than automating the full ISD process. The core design decisions and iterative refinements still require human expertise.

\begin{figure*}[t]
\centering
\includegraphics[width=0.8\textwidth]{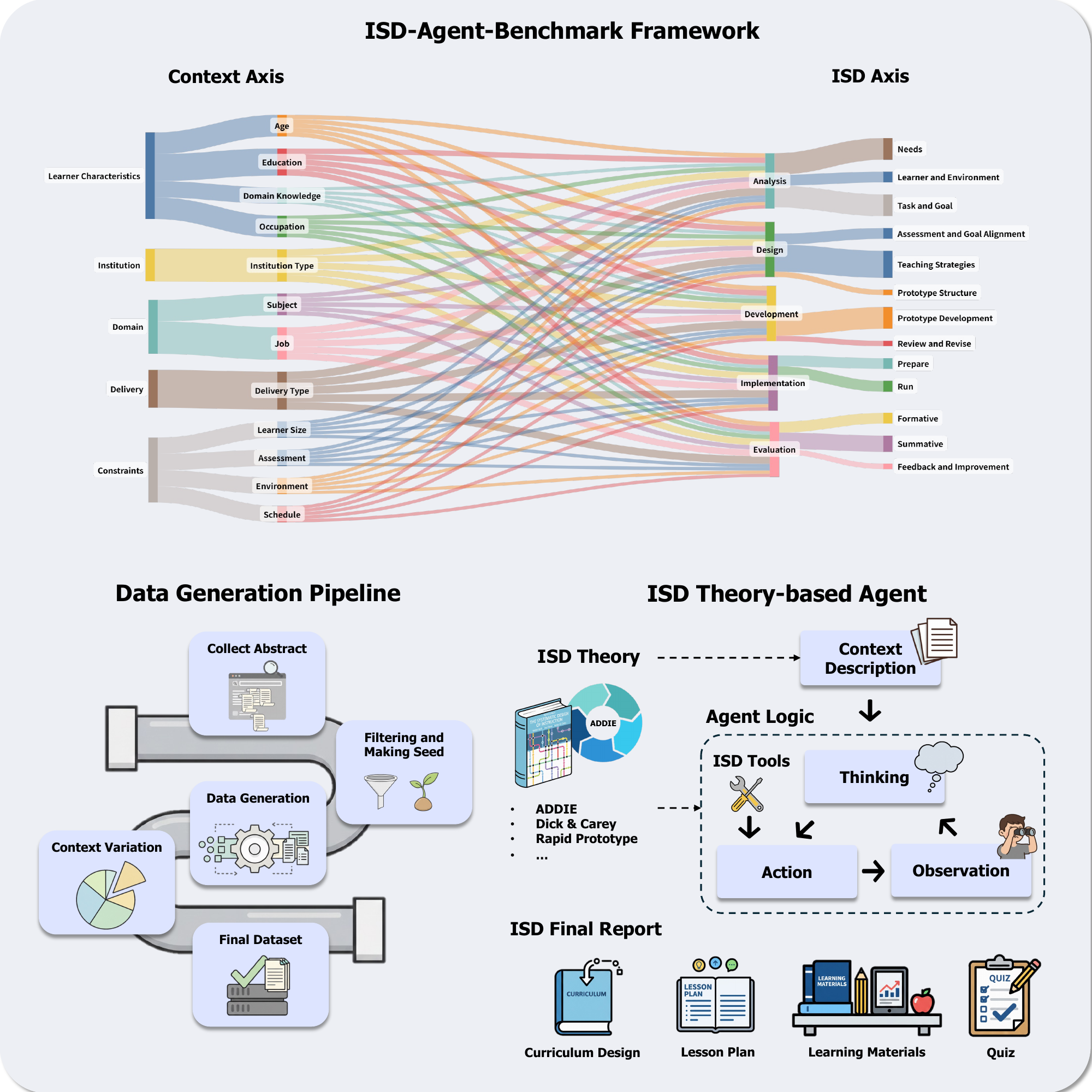}
\caption{Overview of ISD-Agent-Bench. \textit{Top} is the Context Matrix framework that combines 51 contextual variables across 5 categories (Learner Characteristics, Institution, Domain, Delivery, Constraints) with the ISD Matrix containing 33 ADDIE-based sub-steps for systematic scenario generation. \textit{Bottom left} is the data generation pipeline from abstract collection through context variation to final dataset. \textit{Bottom right} is the ISD theory-based agent architecture integrating classical ISD theories with ReAct-style reasoning to produce comprehensive ISD outputs.}
\label{fig:teaser}
\end{figure*}

Meanwhile, LLM-based agents, systems that use LLMs as central controllers to perceive environments, formulate plans, utilize tools, and execute tasks autonomously~\cite{wang2024llmagentsurvey}, have advanced rapidly across diverse domains. In software engineering, agents now resolve real-world GitHub issues~\cite{jimenez2024swebench}; in web navigation, they complete complex multi-step tasks autonomously~\cite{zhou2023webarena}. The evaluation of such agents has also matured, with benchmarks like AgentBench~\cite{liu2023agentbench} assessing capabilities across interactive environments. Notably, agent benchmarks are fundamentally more challenging to develop than traditional LLM benchmarks: while LLMs are evaluated on text generation or question answering, agents operate in dynamic, interactive environments requiring reasoning, planning, tool use, and real-world effects~\cite{mohammadi2025agenteval}.

Despite these advances, research on LLM agents for instructional systems design remains scarce. EduPlanner~\cite{zhang2025eduplanner} applies multi-agent systems to mathematics lesson design, and Instructional Agents~\cite{yao2025instructionalagents} automates course material generation. However, no comprehensive benchmark exists for systematically evaluating ISD agents. Existing educational benchmarks such as EduBench~\cite{edubench2025} and TutorBench~\cite{tutorbench2025} focus on direct LLM evaluation for tutoring or content knowledge, rather than agentic systems addressing the full ISD lifecycle.

To address this gap, we present \benchmark{}\footnote{\url{https://anonymous.4open.science/r/isd-agent-benchmark-8D77}}, the first comprehensive benchmark for evaluating LLM-based ISD agents (Figure~\ref{fig:teaser}). Our benchmark introduces a Context Matrix framework that systematically combines 51 contextual variables with 33 ISD sub-steps derived from the ADDIE model, generating \numscenarios{} diverse scenarios with \numtest{} test cases. The evaluation methodology combines ADDIE-based rubric assessment with trajectory analysis, enabling holistic evaluation of both output quality and agent behavior.

Furthermore, we propose four ISD theory-based agents: React-ADDIE, which integrates the ADDIE framework with ReAct-style reasoning; ADDIE-Agent with fine-grained phase decomposition; Dick-Carey-Agent based on Dick \& Carey's systematic design~\cite{dick1978systematic}; and RPISD-Agent based on Rapid Prototyping ISD~\cite{tripp1990rapid}. Evaluating across three LLMs (Gemini-3-Flash, GPT-5-mini, Solar-Pro3) with a multi-judge protocol achieving 0.905 reliability, our experiments reveal that React-ADDIE achieves the highest performance (86.49 points), demonstrating that the synergy between classical ISD theory and modern agent architectures produces superior outcomes compared to either pure theory-based approaches (ADDIE-Agent: 82.96) or technique-only baselines (84.07).

\subsection{Contributions}
\label{subsec:contributions}

Our contributions are fourfold:

\begin{itemize}
    \item \textbf{Context Matrix Framework}: We propose a two-dimensional framework combining 51 contextual variables across 5 categories with 33 ISD sub-steps for systematic scenario generation, enabling comprehensive coverage of the ISD space.

    \item \textbf{Large-scale Benchmark}: We construct \benchmark{} with \numscenarios{} scenarios and \numtest{} test cases, featuring stratified sampling across 9 contextual dimensions including learner characteristics, institutional context, and difficulty levels.

    \item \textbf{ISD Theory-based Agents}: We develop four agents grounded in classical ISD theories (React-ADDIE, ADDIE-Agent, Dick-Carey-Agent, RPISD-Agent), demonstrating that integrating traditional ISD frameworks with modern agent architectures yields superior performance.

    \item \textbf{Comprehensive Analysis}: We conduct extensive experiments revealing that ISD theory functions as quality assurance rather than performance optimization, with theory value amplifying as task complexity increases.
\end{itemize}

\section{Related Work}
\label{sec:related-work}

\subsection{Agent Benchmark}
\label{subsec:agent-benchmarks}

LLM-based agent evaluation has advanced rapidly with benchmarks like AgentBench~\cite{liu2023agentbench} for diverse interactive environments, WebArena~\cite{zhou2023webarena} for web navigation, and SWE-bench~\cite{jimenez2024swebench} for software engineering. Berkeley Function-Calling Leaderboard~\cite{yan2024bfcl} and ToolEmu~\cite{ruan2024toolemu} further evaluate tool use and safety considerations.

However, educational agent benchmarks remain limited. Existing benchmarks like OpenLearnLM~\cite{openlearnlm2026}, EduBench~\cite{edubench2025}, MathTutorBench~\cite{mathtutorbench2025}, and TutorBench~\cite{tutorbench2025} evaluate LLMs directly rather than agentic systems. The systematic evaluation of agents for ISD, requiring multi-step planning, contextual adaptation, and iterative refinement, remains unexplored.

\subsection{Instructional Systems Design and LLM Agents}
\label{subsec:isd-benchmarks}

ISD is a systematic and reflective process for developing instructional materials and experiences. Unlike narrower instructional design (ID) approaches that focus on specific learning tasks, ISD addresses the entire lifecycle of educational program development, from initial needs analysis through implementation and evaluation. The field emerged from military training research in the 1970s and has since become foundational across corporate training, higher education, and K-12 contexts.

Three theoretical frameworks dominate ISD practice. The ADDIE model~\cite{branson1975addie} remains the most widely adopted framework. Dick \& Carey's systematic design~\cite{dick1978systematic} emphasizes alignment between learning objectives, assessments, and instruction through a nine-step procedure. Rapid Prototyping ISD~\cite{tripp1990rapid} prioritizes iterative development over comprehensive upfront analysis.

Recent work has begun applying LLMs to ISD tasks. EduPlanner~\cite{zhang2025eduplanner} employs multi-agent systems for mathematics lesson design, LessonPlanner~\cite{lessonplanner2024} assists teachers in preparing lesson plans, and LearnLM~\cite{learnlm2024} incorporates learning science principles. However, these systems are often limited to specific domains or do not address the full ISD lifecycle.

Table~\ref{tab:benchmark-comparison} compares existing approaches. Two gaps emerge: (1) current benchmarks focus on content knowledge or tutoring skills rather than the full ISD lifecycle, and (2) existing systems lack agents grounded in established ISD theories. \benchmark{} addresses these gaps by providing both a comprehensive benchmark with systematic scenario generation and novel ISD theory-based agents for evaluation.

\begin{table*}[!t]
\centering
\small
\caption{Comparison of \benchmark{} with existing educational AI benchmarks and systems. \checkmark: fully supported, \textcircled{\small P}: partially supported, --: not supported.}
\label{tab:benchmark-comparison}
\begin{tabular}{lcccccccc}
\toprule
\textbf{Benchmark/System} & \textbf{Focus} & \textbf{\# Items} & \textbf{Domain} & \textbf{\# Dims} & \textbf{Public} & \textbf{Agent} & \textbf{ISD} & \textbf{Evaluation} \\
\midrule
OpenLearnLM~\cite{openlearnlm2026} & Knowledge/Skill/Attitude & 124K & Multi & 3 & \checkmark & -- & \textcircled{\small P} & KSA Rubric \\
EduBench~\cite{edubench2025} & Educational Scenarios & 18.8K & Multi & 9 & \checkmark & -- & -- & Multi-metric \\
MathTutorBench~\cite{mathtutorbench2025} & Math Tutoring & 4.8K & Math & 1 & \checkmark & -- & -- & Pedagogical \\
TutorBench~\cite{tutorbench2025} & Tutoring Skills & 1,490 & Math & 2 & \checkmark & -- & -- & Rubric-based \\
LearnLM~\cite{learnlm2024} & Learning Science & -- & General & -- & -- & -- & \textcircled{\small P} & Principle-based \\
EduPlanner~\cite{zhang2025eduplanner} & Lesson Planning & -- & Math & 1 & -- & \checkmark & \textcircled{\small P} & CIDDP \\
LessonPlanner~\cite{lessonplanner2024} & Lesson Planning & -- & General & -- & \checkmark & -- & \textcircled{\small P} & User study \\
\midrule
\textbf{\benchmark{} (Ours)} & \textbf{ISD} & \textbf{\numscenarios{}} & \textbf{Multi} & \textbf{51} & \checkmark & \checkmark & \checkmark & \textbf{ADDIE} \\
\bottomrule
\end{tabular}
\end{table*}

\section{\benchmark{}}
\label{sec:benchmark}

This section presents our benchmark, including the Context Matrix framework for systematic scenario generation and the evaluation methodology.

\subsection{Context Matrix Framework}
\label{subsec:context-matrix}

The core novelty of our work lies in the Context Matrix, a two-dimensional framework designed to systematically generate comprehensive ISD scenarios. The framework is grounded in a systematic literature review of ISD research, synthesizing contextual factors identified across ISD studies~\cite{branch2009addie,dick1978systematic,morrison2019designing}. The framework combines a Context Axis capturing educational settings with an ISD Axis defining evaluation criteria. Unlike existing educational benchmarks that rely on manually curated scenarios or domain-specific datasets, our framework enables systematic coverage of the ISD space through principled combination of contextual variables.


\subsubsection{Context Axis}
The Context Axis comprises 51 items organized into five categories that collectively define the educational setting for each scenario. Table~\ref{tab:context-axis} summarizes the categories and their item counts.

\begin{table}[t]
\centering
\small
\caption{Context Axis categories and items.}
\label{tab:context-axis}
\begin{tabular}{lcp{4.5cm}}
\toprule
\textbf{Category} & \textbf{\#} & \textbf{Examples} \\
\midrule
Learner Characteristics & 16 & Age groups, education levels, expertise, occupation \\
Institutional Context & 6 & K-12, university, corporate, vocational \\
Educational Domain & 10 & Language, math, science, IT, healthcare \\
Delivery Mode & 7 & Classroom, online, blended, VR/simulation \\
Constraints & 12 & Class size, duration, technology, assessment \\
\midrule
\textbf{Total} & \textbf{51} & \\
\bottomrule
\end{tabular}
\end{table}

Learner Characteristics capture age groups ranging from teens to adults over 40, education levels from elementary to adult non-degree learners, domain expertise levels, and occupational roles including students, office workers, professionals, and teachers. Institutional Context covers the spectrum of educational institutions from K-12 schools through universities and graduate programs to corporate training and vocational institutions. Educational Domain spans both academic subjects such as language, mathematics, science, and social studies, as well as vocational areas including software development, AI, healthcare, business, and customer service. Delivery Mode encompasses traditional classroom instruction, synchronous and asynchronous online learning, blended approaches, mobile microlearning, simulation/VR-based instruction, and project-based learning. Constraints address practical considerations including class size, assessment requirements, technology availability, and program duration.

\subsubsection{ISD Axis}
The ISD Axis defines 33 sub-steps derived from the ADDIE model, organized into 13 aggregated evaluation items across the five phases. The Analysis phase encompasses needs analysis (problem identification, gap analysis, performance analysis, needs prioritization), learner and environment analysis, and task and goal analysis including learning objectives, subordinate skills, and entry behaviors. The Design phase covers assessment and goal alignment design, instructional strategy and learning experience design, and prototype structure design. The Development phase includes prototype development for learner materials, instructor manuals, administrator guides, and assessment tools, followed by expert review. The Implementation phase addresses program preparation through orientations and system checks, as well as program execution and monitoring. The Evaluation phase comprises formative evaluation with pilot testing and revision, summative evaluation with effectiveness analysis and adoption decisions, and program improvement feedback loops.

\subsection{Scenario Generation Pipeline}
\label{subsec:scenario-generation}

Our scenario generation pipeline systematically combines Context Axis variables to create diverse, realistic ISD challenges through five stages. First, we collected 10,577 paper abstracts from SCOPUS-indexed journals using the query ``instructional design'' OR ``learning design'' in educational technology venues. After filtering for abstract length ($\geq$200 characters), removing duplicates by title and DOI, and requiring author keywords, 8,842 papers remained as seed contexts. These papers exhibit domain concentration: language, science, mathematics, and social studies each comprise approximately 24--25\%, while education-specific (1.9\%) and healthcare (0.1\%) domains are underrepresented.

Second, we sample from the cross-product of learner characteristics, institutional context, domain, delivery mode, and constraints to ensure comprehensive coverage of the scenario space. Given the combinatorial explosion of possible combinations ($51^5$ theoretical possibilities), we employ stratified sampling to achieve balanced representation across all categories. Third, for each seed paper, we use GPT-4o to generate scenario content aligned with the extracted context, including educational goals in SMART format, contextual constraints and requirements, prior knowledge specifications, and available resources and limitations. This produces 8,842 seed-based scenarios. To address domain imbalance and context gaps, we generate 16,953 additional scenarios through context variation and targeted synthesis for underrepresented combinations.

Fourth, quality control proceeds through two validation stages: rule-based validation ensures completeness of required fields and consistency of constraint combinations, while LLM-based review checks for logical consistency between context elements and scenario content, filtering out implausible combinations. Finally, to address natural imbalances in context distributions, we employed targeted augmentation strategies, generating additional scenarios for underrepresented categories including teenage learners, large class sizes (30+ students), and adult self-directed learning contexts.

Difficulty levels are computed from a weighted composite of scenario characteristics: number of learning goals (weight 0.25), domain expertise level (0.25), number of resources (0.20), course duration (0.20), and budget constraints (0.10). Easy scenarios typically have $\leq$3 learning goals, beginner-level expertise, and short-term duration; Hard scenarios have $\geq$4 learning goals, advanced expertise, and long-term duration. Scenarios are ranked by composite score and assigned to Easy, Medium, or Hard with target distribution 33:34:33 to ensure balanced evaluation. The following box shows a representative Hard scenario; additional examples are provided in Appendix~\ref{app:examples}.

\begin{tcolorbox}[colback=gray!5!white, colframe=gray!75!black, title=\small\textbf{Example Scenario (Hard)}]
\small
\textbf{Task:} Design a game-based business leadership course for corporate professionals (30--40s) in online synchronous environment. \\
\textbf{Learners:} Mid-level managers with intermediate leadership experience, class size 20--30. \\
\textbf{Constraints:} Gamification required, real-time interaction, BYOD, project-based assessment, 2--4 weeks duration. \\
\textbf{Objectives:} Develop strategic decision-making and cross-functional collaboration skills. \\
\textbf{Evaluation:} 33 ADDIE sub-steps across 5 phases (Analysis, Design, Development, Implementation, Evaluation).
\end{tcolorbox}

\subsection{Dataset Statistics}
\label{subsec:dataset-stats}

The final dataset comprises \numscenarios{} scenarios split into training (24,593; 95.3\%) and test (\numtest{}; 4.7\%) sets. Table~\ref{tab:dataset-distribution} presents the distribution across key contextual dimensions. The stratified sampling procedure ensures balanced representation, with learner age ranging from 16.8\% (teens) to 32.4\% (20s), and difficulty levels evenly distributed around 33\% each. The maximum distribution difference between train and test sets remains below 3.54\% for any dimension.

\begin{table}[t]
\centering
\small
\caption{Test dataset distribution across key dimensions (n=\numtest{}).}
\label{tab:dataset-distribution}
\begin{tabular}{llrr}
\toprule
\textbf{Dimension} & \textbf{Value} & \textbf{Count} & \textbf{\%} \\
\midrule
\multirow{4}{*}{Learner Age}
    & Teens (13-19) & 202 & 16.8 \\
    & In their 20s & 390 & 32.4 \\
    & In their 30s & 286 & 23.8 \\
    & 40s and above & 324 & 27.0 \\
\midrule
\multirow{3}{*}{Class Size}
    & Small (1-10) & 335 & 27.9 \\
    & Medium (10-30) & 289 & 24.0 \\
    & Large (30+) & 578 & 48.1 \\
\midrule
\multirow{3}{*}{Difficulty}
    & Easy & 397 & 33.0 \\
    & Moderate & 405 & 33.7 \\
    & Hard & 400 & 33.3 \\
\bottomrule
\end{tabular}
\end{table}

\subsection{Evaluation Methodology}
\label{subsec:evaluation}

Our evaluation employs a two-stage LLM-as-a-Judge approach using GPT-4o, combining ADDIE rubric assessment (70\%) with trajectory evaluation (30\%).

\subsubsection{ADDIE Rubric Evaluation}
The ADDIE rubric evaluates output quality across 13 aggregated items mapped from 33 sub-steps. Each item is scored on a 5-level scale from Absent (1-2 points) through Poor (3-4), Satisfactory (5-6), and Good (7-8) to Excellent (9-10). The evaluation proceeds in two stages: first, a status assessment determines whether each component is absent, weak, moderate, good, or excellent; second, numerical scores are assigned within the appropriate range based on the status determination.

Phase weights reflect the relative importance of each ADDIE component: Analysis and Design each receive 25\%, Development receives 20\%, and Implementation and Evaluation each receive 15\%. These weights can be dynamically adjusted based on scenario context, with Development weighted higher for teenage learners or simulation-based delivery, and Analysis weighted higher for adult learners or advanced expertise levels.

\subsubsection{Trajectory Evaluation}
Following the Berkeley Function-Calling Leaderboard methodology, trajectory evaluation assesses the agent's tool usage patterns across four dimensions. Tool Correctness (25 points) measures whether appropriate tools were selected for each task. Argument Accuracy (25 points) evaluates the correctness of parameters passed to tools. Redundancy Avoidance (25 points) penalizes unnecessary or duplicate tool calls. Result Utilization (25 points) assesses how effectively tool outputs were incorporated into the final design.

The final score combines both components: $\text{Score} = 0.7 \times \text{ADDIE} + 0.3 \times \text{Trajectory}$.

\begin{table*}[!t]
\centering
\small
\caption{Comparison of baseline and theory-based agents on 1,017 test scenarios. React-ADDIE, combining ISD theory with ReAct architecture, achieves the best performance. Best in \textbf{bold}.}
\label{tab:main-results}
\begin{tabular}{l cccc ccc ccccc}
\toprule
& \multicolumn{4}{c}{\textbf{Evaluator Model}} & \multicolumn{3}{c}{\textbf{Difficulty}} & \multicolumn{5}{c}{\textbf{Phase-wise (3-model avg)}} \\
\cmidrule(lr){2-5} \cmidrule(lr){6-8} \cmidrule(lr){9-13}
\textbf{Agent} & \textbf{Gemini} & \textbf{GPT} & \textbf{Solar} & \textbf{Avg} & \textbf{Easy} & \textbf{Med.} & \textbf{Hard} & \textbf{Ana.} & \textbf{Des.} & \textbf{Dev.} & \textbf{Impl.} & \textbf{Eval.} \\
\midrule
\multicolumn{13}{l}{\textit{Comparison}} \\
\quad Baseline & 83.41 & 85.01 & 83.78 & 84.07 & 83.92 & 84.15 & 84.12 & 81.5 & 80.8 & 83.2 & 81.0 & 83.5 \\
\quad EduPlanner~\cite{zhang2025eduplanner} & 83.23 & 61.43 & 61.46 & 68.70 & 69.12 & 68.54 & 68.45 & 68.2 & 65.4 & 48.5 & 62.8 & 58.0 \\
\midrule
\multicolumn{13}{l}{\textit{Theory-based (Ours)}} \\
\quad ADDIE-Agent & 83.18 & 84.25 & 81.43 & 82.96 & 82.78 & 83.04 & 83.05 & 79.5 & 78.2 & 77.8 & 80.8 & 79.2 \\
\quad RPISD-Agent & 83.35 & 85.59 & 81.93 & 83.62 & 83.41 & 83.68 & 83.77 & 81.2 & 80.5 & 79.8 & 83.2 & 82.5 \\
\quad Dick-Carey-Agent & 81.13 & 87.92 & 83.55 & 84.20 & 83.98 & 84.27 & 84.35 & 80.8 & 79.5 & \textbf{84.2} & 81.5 & 80.8 \\
\quad React-ADDIE & \textbf{87.61} & \textbf{88.43} & \textbf{83.44} & \textbf{86.49} & \textbf{86.21} & \textbf{86.52} & \textbf{86.73} & \textbf{84.2} & \textbf{84.8} & 82.5 & \textbf{84.0} & \textbf{83.0} \\
\bottomrule
\end{tabular}
\end{table*}

\section{ISD Theory-based Agents}
\label{sec:agents}

This section formalizes the agent architectures evaluated in our benchmark. We first define the general agent framework, then describe four ISD theory-based agents that integrate classical ISD theories with modern LLM agent architectures.

\subsection{Problem Formulation}
\label{subsec:formulation}

Given an ISD scenario $s \in \mathcal{S}$ containing learner characteristics, institutional context, educational domain, delivery mode, and constraints, an agent $\mathcal{A}$ generates an ISD output $o$ through a sequence of actions:
\begin{equation}
o = \mathcal{A}(s) = f_{\text{gen}}(s, \{(a_1, r_1), (a_2, r_2), \ldots, (a_T, r_T)\})
\end{equation}
where $a_t$ denotes an action at step $t$, $r_t$ is the corresponding result, and $T$ is the total number of steps. The output $o$ is evaluated against ADDIE criteria to produce a score $\text{Score}(o) \in [0, 100]$.

\subsection{ADDIE-Agent}
\label{subsec:addie-agent}

ADDIE-Agent implements the classical five-phase sequential workflow~\cite{branson1975addie} with fine-grained tool decomposition. The agent uses 14 specialized tools that break down each ADDIE phase into discrete sub-tasks (e.g., \texttt{analyze\_needs}, \texttt{analyze\_learner}, \texttt{analyze\_context}, \texttt{design\_objectives}, \texttt{design\_assessment}).

Each phase $p \in \{A, D, D', I, E\}$ (Analysis, Design, Development, Implementation, Evaluation) produces an intermediate output $o_p$ through multiple tool calls that condition subsequent phases:
\begin{equation}
o_p = f_p(s, \{o_{p'} : p' \prec p\}) = \bigcup_{t \in \mathcal{T}_p} \text{Execute}(t, s, h)
\end{equation}
where $\mathcal{T}_p$ denotes the tool subset for phase $p$ and $p' \prec p$ denotes phases preceding $p$ in the ADDIE sequence. Phase transitions occur only after validation:
\begin{equation}
\text{Proceed}(p \to p') = \mathbb{1}[\text{Valid}(o_p) = \text{True}]
\end{equation}
This fine-grained decomposition provides detailed phase outputs but incurs higher computational overhead.

\subsection{Dick-Carey-Agent}
\label{subsec:dick-carey-agent}

Dick-Carey-Agent follows the nine-step systematic design model~\cite{dick1978systematic}, emphasizing alignment between objectives, assessments, and instruction. The agent executes steps $\{s_1, \ldots, s_9\}$ with explicit dependency tracking:
\begin{equation}
o_{s_i} = f_{s_i}(s, \{o_{s_j} : s_j \in \text{Deps}(s_i)\})
\end{equation}
where $\text{Deps}(s_i)$ returns prerequisite steps for $s_i$. Key steps include:
\begin{itemize}
    \item $s_1$: Identify instructional goals
    \item $s_2$: Conduct instructional analysis
    \item $s_3$: Analyze learners and contexts
    \item $s_4$: Write performance objectives
    \item $s_5$: Develop assessment instruments
    \item $s_6$: Develop instructional strategy
    \item $s_7$: Develop and select instructional materials
    \item $s_8$: Design and conduct formative evaluation
    \item $s_9$: Revise instruction
\end{itemize}
The model enforces criterion-referenced assessment by requiring $s_5$ (assessment) to directly derive from $s_4$ (objectives).

\subsection{RPISD-Agent}
\label{subsec:rpisd-agent}

RPISD-Agent implements Rapid Prototyping ISD~\cite{tripp1990rapid}, prioritizing early prototype generation and iterative refinement over comprehensive upfront analysis. The agent operates in $K$ cycles:
\begin{equation}
o^{(k)} = f_{\text{refine}}(o^{(k-1)}, \text{Feedback}(o^{(k-1)})), \quad k = 1, \ldots, K
\end{equation}
where $o^{(0)}$ is an initial rapid prototype generated with minimal analysis, and $\text{Feedback}(\cdot)$ simulates stakeholder evaluation. Each cycle applies targeted improvements:
\begin{equation}
\Delta o^{(k)} = \text{LLM}(\text{Critique}(o^{(k-1)}), s)
\end{equation}
We set $K=3$ cycles, balancing iteration depth with computational efficiency. This approach is optimized for time-constrained scenarios where rapid delivery outweighs exhaustive planning.

\subsection{React-ADDIE Agent}
\label{subsec:react-addie}

React-ADDIE integrates the classical ADDIE framework with ReAct-style reasoning~\cite{yao2023react}, implementing each ISD phase as a single integrated tool call. Unlike ADDIE-Agent's fine-grained 14-tool approach, React-ADDIE uses 5 phase-level tools $\mathcal{T} = \{t_A, t_D, t_{D'}, t_I, t_E\}$ corresponding to Analysis, Design, Development, Implementation, and Evaluation:
\begin{equation}
o_p = \text{ReAct}(s, h_{p-1}, t_p), \quad p \in \{A, D, D', I, E\}
\end{equation}
where each tool $t_p$ encapsulates the complete reasoning and action for an entire ADDIE phase within a single LLM call. At each phase, the agent generates a thought $\tau_p$, executes the phase tool $t_p$, and observes the result $r_p$:
\begin{equation}
(\tau_p, o_p) = \text{LLM}(s, h_{p-1}, t_p), \quad h_p = h_{p-1} \cup \{(\tau_p, t_p, o_p)\}
\end{equation}
This design reduces multi-step decomposition overhead while maintaining theoretical grounding in the ADDIE framework, enabling efficient yet systematic ISD.

\section{Experiments}
\label{sec:experiments}

\subsection{Experimental Setting}
\label{subsec:setting}

We evaluate six agents on 1,017 test scenarios from \benchmark{} using three base language models: Gemini-3-Flash, GPT-5-mini, and Solar-Pro3. All models use temperature 0.7 with a maximum of 10 interaction turns per scenario.

Evaluation employs a multi-judge protocol using two LLMs from different providers: GPT-4o-mini (OpenAI) and Gemini-2.5-flash-lite (Google). This cross-provider design mitigates potential self-preference bias where models may favor outputs from the same provider. Final scores are computed using median aggregation across judges to ensure robustness.

\begin{figure*}[!t]
\centering
\includegraphics[width=\textwidth]{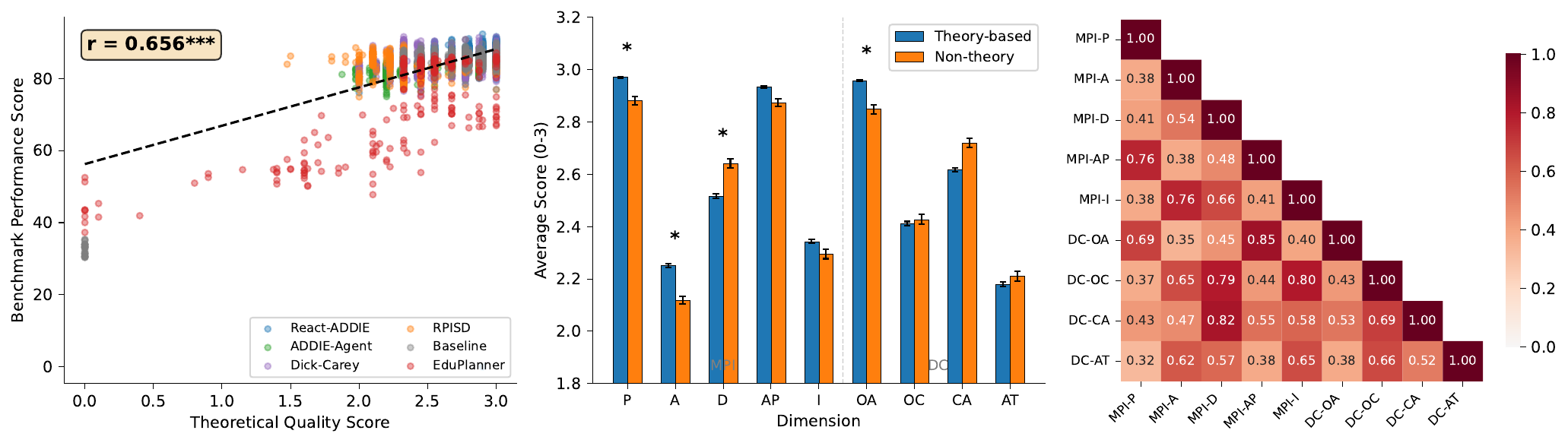}
\caption{Theoretical quality analysis. \textit{Left panel} shows correlation between theoretical quality scores and benchmark performance ($r = 0.656$, $p < 0.001$). \textit{Middle panel} compares theory-based and non-theory agents across all 9 dimensions, where asterisks indicate significant differences ($d > 0.2$). \textit{Right panel} displays inter-dimension correlation matrix revealing cross-framework relationships.}
\label{fig:findings-1-3}
\end{figure*}

\subsection{Overall Performance}
\label{subsec:overall-performance}

Table~\ref{tab:main-results} presents performance across all agents on 1,017 test scenarios evaluated with three LLMs. React-ADDIE achieves the highest mean score (86.49), demonstrating that integrating the ADDIE framework with ReAct-style reasoning yields superior ISD quality.

React-ADDIE consistently achieves the highest scores across all three LLMs, with particularly strong performance on Gemini-3-Flash (87.61) and GPT-5-mini (88.43). The phase-wise analysis reveals that React-ADDIE leads in Analysis (84.2), Design (84.8), Implementation (84.0), and Evaluation (83.0), while Dick-Carey-Agent excels specifically in Development (84.2) due to its systematic materials creation approach.

A notable observation is EduPlanner's performance degradation on GPT-5-mini (61.43) and Solar-Pro3 (61.46), with particularly weak Development phase scores (48.5). This suggests that EduPlanner's multi-agent architecture may be sensitive to model-specific characteristics, while our theory-based agents show more consistent cross-model performance.

React-ADDIE's integration of ADDIE theory with ReAct-style reasoning outperforms both pure theory-based agents (ADDIE-Agent: 82.96, which uses the same ADDIE framework but with fine-grained 14-tool decomposition) and technique-only approaches (Baseline: 84.07). This suggests that the optimal approach lies in combining established ISD theories with modern agent architectures.

\subsection{Multi-Judge Evaluation}
\label{subsec:multi-judge}

To address LLM-as-a-Judge bias concerns, we employ two judge models from different providers: GPT-4o-mini (OpenAI) and Gemini-2.5-flash-lite (Google). Table~\ref{tab:reliability} summarizes inter-judge agreement and bias metrics.

\begin{table}[t]
\centering
\small
\caption{Inter-judge reliability and bias analysis across 5,183 evaluations.}
\label{tab:reliability}
\begin{tabular}{llc}
\toprule
\textbf{Category} & \textbf{Metric} & \textbf{Value} \\
\midrule
\multirow{3}{*}{Agreement} & Mean Reliability & 0.905 \\
& Std. Deviation & 0.084 \\
& Min / Max & 0.293 / 1.000 \\
\midrule
\multirow{4}{*}{Distribution} & Excellent ($\geq$0.9) & 63.3\% \\
& Good (0.75--0.9) & 31.4\% \\
& Moderate (0.5--0.75) & 5.1\% \\
& Low ($<$0.5) & 0.3\% \\
\midrule
\multirow{2}{*}{Bias} & Gemini-2.5-flash-lite & +0.060 \\
& GPT-4o-mini & -0.060 \\
\bottomrule
\end{tabular}
\end{table}

The high inter-judge reliability (0.905) with 94.7\% of evaluations achieving good or excellent agreement validates our multi-judge protocol. Both judges show minimal systematic bias ($\pm$0.06 points from median), confirming balanced evaluation without provider preference.

\begin{figure*}[!t]
\centering
\includegraphics[width=\textwidth]{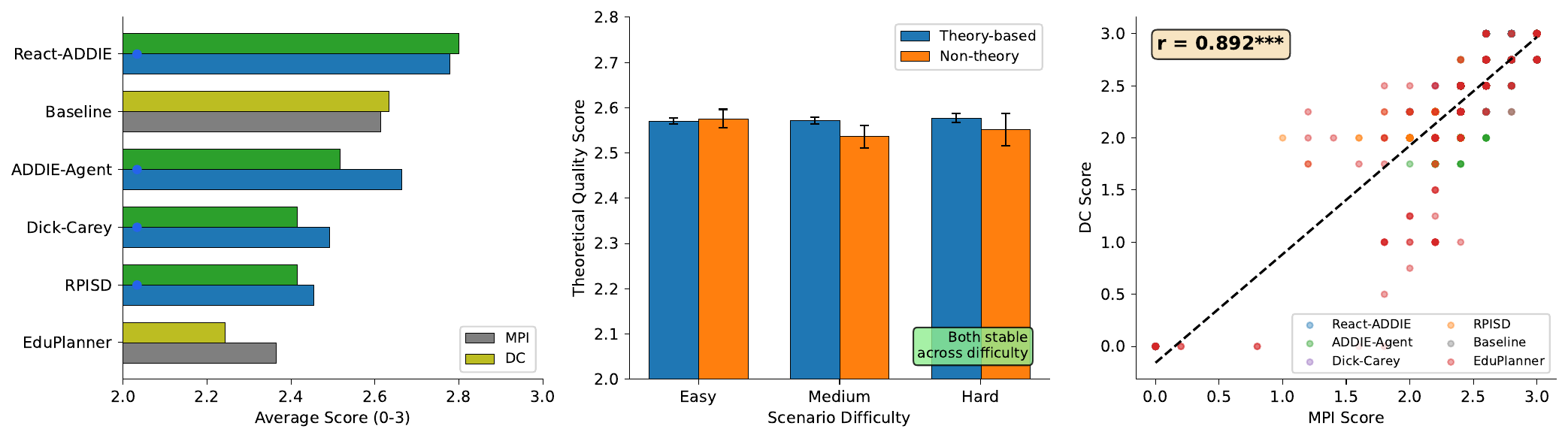}
\caption{Agent comparison and framework validation. \textit{Left panel} presents agent-wise MPI and DC scores with theory-based agents marked, where React-ADDIE leads both dimensions. \textit{Middle panel} shows theoretical quality by difficulty level, demonstrating both agent types maintain stable performance. \textit{Right panel} illustrates MPI-DC correlation ($r = 0.892$), confirming the frameworks capture related but complementary quality aspects.}
\label{fig:findings-4-6}
\end{figure*}

\section{Analysis}
\label{sec:analysis}

Beyond benchmark performance scores, we analyze the theoretical quality of agent outputs using two established ISD frameworks: Merrill's First Principles of Instruction \cite{merrill2002first} (MPI) and Design Coherence (DC), which extends constructive alignment \cite{biggs1996enhancing} to include technology integration. We labeled 5,345 outputs across 9 dimensions (MPI: 5, DC: 4) using GPT-4o-mini with strict rubrics (0-3 scale). This analysis compares theory-based agents (React-ADDIE, ADDIE-Agent, Dick-Carey, RPISD) against non-theory agents (Baseline, EduPlanner).

\subsection{Theoretical Quality Predicts Performance}

As shown in Figure~\ref{fig:findings-1-3} \textit{left}, we find a strong positive correlation between theoretical quality scores and benchmark performance ($r = 0.656$, $p < 0.001$), validating that adherence to ISD principles translates to measurable quality improvements. This correlation holds across all agents, with EduPlanner showing the highest within-agent correlation ($r = 0.312$), suggesting its failures are systematically related to theoretical deficiencies rather than random variation.

The correlation validates two claims: (1) our MPI/DC labeling captures meaningful quality dimensions, and (2) the benchmark effectively measures theoretically-grounded ISD quality. Practitioners can use theoretical quality scores as diagnostic tools to identify specific improvement areas.

\subsection{Theory-based vs Non-theory Agent Comparison}

Figure~\ref{fig:findings-1-3} \textit{middle} compares the two agent groups across all 9 dimensions. Across MPI dimensions, theory-based agents significantly outperform non-theory agents on Problem-centered design (P: $d = 0.28$) and Activation of prior knowledge (A: $d = 0.29$), both $p < 0.001$. These dimensions require explicit consideration of learner context and knowledge scaffolding---precisely where theoretical frameworks provide structured guidance. In contrast, non-theory agents score higher on Demonstration (D: $d = -0.23$, $p < 0.001$), suggesting a tradeoff between systematic coverage and depth in concrete examples.

For DC dimensions, theory-based agents demonstrate significantly better Objective-Assessment alignment (OA: $d = 0.32$, $p < 0.001$), reflecting systematic mapping between learning goals and evaluation. However, Content-Activity alignment (CA) favors non-theory agents ($d = -0.19$, $p < 0.001$), paralleling the MPI-D finding: simpler agents may produce more tightly coupled content-activity pairs. Activity-Technology alignment (AT) remains the weakest dimension across all agents (mean = 2.19), identifying technology integration as a universal challenge.

\subsection{Cross-Framework Dimension Relationships}

The correlation matrix in Figure~\ref{fig:findings-1-3} \textit{right} reveals meaningful pathways between MPI and DC dimensions. The strongest cross-framework correlation appears between Demonstration and Content-Activity alignment ($r = 0.82$), indicating that agents excelling at concrete examples naturally produce activities matching their content. Similarly, Application correlates strongly with Objective-Assessment alignment ($r = 0.85$), suggesting that practice opportunities and valid assessment design emerge together. Integration shows substantial correlation with Objective-Content alignment ($r = 0.80$), where real-world transfer emphasis accompanies coherent goal-material mapping.

These pathways suggest that MPI and DC form an interconnected network rather than independent quality dimensions. Improving one dimension may cascade to related dimensions, offering leverage points for agent optimization. For instance, enhancing demonstration capabilities could simultaneously improve content-activity coherence.

\subsection{Agent-wise Theoretical Quality}

Figure~\ref{fig:findings-4-6} \textit{left} presents agent-wise theoretical quality scores. React-ADDIE achieves the highest scores on both MPI (2.78) and DC (2.80), consistent with its benchmark performance leadership. The ranking largely follows benchmark results, with one notable exception: Baseline (MPI: 2.61, DC: 2.63) outperforms several theory-based agents in DC despite its simpler design.

EduPlanner shows the lowest theoretical quality (MPI: 2.36, DC: 2.24), explaining its poor benchmark performance. Its multi-agent architecture, while innovative, fails to ensure systematic coverage of ISD principles---components may optimize locally without maintaining global coherence.

The strong correlation between agent rankings on theoretical quality and benchmark performance ($\rho = 0.94$) reinforces that ISD theory adherence is a primary driver of output quality.

\subsection{Robustness Across Difficulty Levels}

As shown in Figure~\ref{fig:findings-4-6} \textit{middle}, both theory-based and non-theory agents maintain stable theoretical quality across Easy, Medium, and Hard scenarios. Theory-based agents show consistent scores (Easy: 2.58, Medium: 2.58, Hard: 2.58), as do non-theory agents (Easy: 2.55, Medium: 2.54, Hard: 2.53).

This stability indicates that agent quality is determined by architectural and prompting choices rather than scenario characteristics. Well-designed agents handle complexity gracefully, while poorly designed agents fail consistently regardless of difficulty. For practitioners, this means agent selection matters more than scenario filtering.

\subsection{MPI and DC as Complementary Frameworks}

Figure~\ref{fig:findings-4-6} \textit{right} shows that MPI and DC total scores correlate strongly ($r = 0.892$, $p < 0.001$), yet the shared variance ($r^2 = 0.80$) leaves 20\% unexplained, confirming they capture related but distinct quality aspects. While cross-framework correlations are substantial Figure~\ref{fig:findings-1-3} \textit{right}, each framework contributes unique diagnostic information.

MPI assesses instructional effectiveness principles, while DC evaluates internal alignment coherence. An agent can demonstrate strong instructional principles (high MPI) while lacking tight component alignment (lower DC), or vice versa. We recommend practitioners evaluate outputs on both dimensions, using MPI for pedagogical soundness and DC for structural coherence.

\section{Conclusion}
\label{sec:conclusion}

We presented \benchmark{}, the first comprehensive benchmark for evaluating LLM-based ISD agents. Our Context Matrix framework combines 51 contextual variables with 33 ADDIE-based sub-steps to generate \numscenarios{} diverse scenarios. React-ADDIE achieves the highest performance (86.49) by integrating classical ISD theories with modern agent architectures. A key finding is that coarse-grained tool design (5 phase-level tools) outperforms fine-grained decomposition (14 tools), suggesting that holistic reasoning with systematic structure yields optimal results.

\subsection{Use of Generative AI}

This paper was written with assistance from Claude-4-Opus and GPT-5.2 for text refinement. All ideas, research design, and intellectual contributions originated from the human authors.



\clearpage
\appendix
\section{Benchmark Scenario Examples}
\label{app:examples}

This section presents representative scenarios from \benchmark{}, illustrating the diversity of instructional contexts and varying performance patterns across different agents. We selected three scenarios representing different difficulty levels (Easy, Medium, Hard) to demonstrate how agent performance varies across complexity levels. Each example includes the complete scenario specification, detailed agent rankings, phase-level performance breakdown, and qualitative analysis of the key differentiating factors between top and bottom performers.

\subsection{Example 1: VR-based Language Learning (Easy)}

The first example represents a relatively straightforward instructional design task involving the development of a mobile-assisted collaborative language learning program in a VR/simulation environment. This scenario was classified as Easy because it involves a well-established pedagogical domain (language learning), targets learners with clear characteristics (university students in their 20s with intermediate proficiency), and imposes minimal operational constraints. The VR/simulation environment, while technologically sophisticated, follows established patterns for immersive language learning that are well-documented in the instructional design literature.

\begin{tcolorbox}[colback=blue!5!white, colframe=blue!50!black, breakable]
\small
Mobile-Assisted Collaborative Language Learning Simulation Program (ID BAL-LARGE2-0075). This scenario targets university students in their 20s who are intermediate-level language learners. The institutional context is a university language department with adequate technology infrastructure. The learning environment is VR/simulation-based, enabling immersive conversational practice with virtual native speakers and peer collaboration in simulated real-world contexts such as restaurants, airports, and business meetings. The program duration is mid-term (2--4 weeks), with a medium class size of 15--25 students. Assessment is formative, focusing on conversational fluency improvement and collaborative task completion. Technology requirements specify that VR headsets and mobile devices are provided by the institution. The primary learning objectives include improving conversational confidence, developing cross-cultural communication skills, and building vocabulary through contextual immersion. Budget constraints are moderate, and the difficulty level is Easy due to simple structure and minimal constraints.
\end{tcolorbox}

Table~\ref{tab:example1} shows the agent rankings for this scenario. Dick-Carey achieves the highest total score (80.3), demonstrating its strength in VR environments through its systematic approach to media selection and detailed learner analysis procedures. The Dick \& Carey model's emphasis on identifying entry behaviors and conducting thorough context analysis proved particularly valuable for designing VR-based instruction, as it ensured the technology choices were grounded in learner characteristics rather than novelty alone.

Notably, the Development phase shows the largest performance variance across agents, with Dick-Carey scoring 81.6 while EduPlanner scores only 45.2. This 36.4-point gap reflects fundamental differences in how agents approach prototype development. Dick-Carey's systematic approach generated detailed storyboards for each VR scenario, complete with branching dialogue trees and specific feedback mechanisms. In contrast, EduPlanner produced generic development plans that failed to address the unique requirements of VR content creation, such as spatial audio design, gesture recognition integration, and motion sickness mitigation strategies.

The Analysis phase scores are notably consistent across all agents (ranging from 74.5 to 75.0), suggesting that the straightforward learner profile and clear institutional context made needs analysis relatively uniform. However, this surface-level similarity masks important qualitative differences. Dick-Carey's analysis included detailed prerequisite skill mapping for VR interaction, while lower-performing agents treated VR as a simple delivery medium rather than an interactive learning environment requiring specific user competencies.

\begin{table}[ht!]
\centering
\small
\caption{Agent rankings for Example 1 (VR Language Learning)}
\label{tab:example1}
\begin{tabular}{clcccc}
\toprule
\textbf{Rank} & \textbf{Agent} & \textbf{Total} & \textbf{ADDIE} & \textbf{Analysis} & \textbf{Develop.} \\
\midrule
1 & Dick-Carey & 80.3 & 77.8 & 74.5 & 81.6 \\
2 & Baseline & 79.3 & 77.3 & 75.0 & 80.9 \\
3 & ReAct-ISD & 79.0 & 76.5 & 75.0 & 77.8 \\
4 & ADDIE & 78.5 & 74.9 & 74.6 & 62.2 \\
5 & EduPlanner & 76.6 & 73.1 & 75.0 & 45.2 \\
6 & RPISD & 72.5 & 67.2 & 74.6 & 58.5 \\
\bottomrule
\end{tabular}
\end{table}

\subsection{Example 2: AI Literacy Assessment (Medium)}

The second example involves designing formative assessment tools for AI literacy education, representing a medium-complexity task that requires balanced attention to both content development and evaluation mechanisms. This scenario presents unique challenges because AI literacy is a rapidly evolving domain where content can become outdated quickly, and assessment must balance technical accuracy with accessibility for non-specialist learners. The medium difficulty classification reflects the need to design assessments that can accurately measure understanding of abstract concepts like algorithmic bias, model interpretability, and ethical AI deployment.

\begin{tcolorbox}[colback=green!5!white, colframe=green!50!black, breakable]
\small
Simulation-based Instructional Design for AI Literacy Formative Assessment (ID BAL-LARGE2-0117). This scenario targets adult learners in their 30s working in corporate settings who need to understand AI concepts for professional decision-making without becoming technical practitioners. The institutional context is a corporate training department seeking to upskill employees across multiple business units. The learning environment combines simulation-based activities with VR components, allowing learners to experience the consequences of AI-driven decisions in safe, consequence-free scenarios such as hiring simulations demonstrating algorithmic bias or customer service chatbot deployment scenarios. Program duration is mid-term (2--4 weeks), with a large class size of 50+ participants distributed across multiple cohorts. Assessment is primarily formative, focusing on developing critical evaluation skills for AI systems rather than technical implementation ability. Technology follows a BYOD (Bring Your Own Device) model with web-based simulations accessible on personal laptops and tablets. Learning objectives include understanding fundamental AI concepts, recognizing limitations and biases in AI systems, making informed decisions about AI adoption in business contexts, and communicating AI-related risks to stakeholders. The difficulty level is Medium due to the moderately complex requirement of balancing technical accuracy with business applicability.
\end{tcolorbox}

As shown in Table~\ref{tab:example2}, Baseline achieves the highest total score (80.0) in this assessment-focused scenario, with particularly strong performance in the Evaluation phase (84.8). This result initially appears counterintuitive, as one might expect theory-grounded agents to outperform the single-shot Baseline. However, closer analysis reveals that the Baseline's comprehensive prompt, which explicitly specifies all 33 ADDIE sub-items including detailed evaluation criteria, provided sufficient structure for this assessment-centric task.

The Baseline's success in this scenario aligns with Finding 5's observation that Evaluation performance often trades off with Implementation. Because this scenario's primary deliverable is formative assessment tools rather than a full instructional program, agents that allocated more attention to evaluation design naturally performed better. The Baseline's holistic approach, which treats all phases with equal attention in a single generation pass, happened to match this scenario's requirements well.

Dick-Carey's second-place finish (79.3) reflects its systematic approach to assessment alignment, though its emphasis on summative evaluation and criterion-referenced testing was less well-suited to the formative, developmental nature of AI literacy assessment. ReAct-ISD (77.3) showed strong Evaluation phase performance (80.9) but lost points in Design due to its iterative approach sometimes producing redundant assessment items across phases. The lower-ranked agents (ADDIE-Agent at 70.9 and RPISD at 70.0) struggled with the abstract nature of AI concepts, producing assessments that tested factual recall rather than critical evaluation skills.

\begin{table}[ht!]
\centering
\small
\caption{Agent rankings for Example 2 (AI Literacy Assessment)}
\label{tab:example2}
\begin{tabular}{clcccc}
\toprule
\textbf{Rank} & \textbf{Agent} & \textbf{Total} & \textbf{ADDIE} & \textbf{Design} & \textbf{Eval.} \\
\midrule
1 & Baseline & 80.0 & 78.3 & 75.5 & 84.8 \\
2 & Dick-Carey & 79.3 & 76.4 & 76.0 & 75.2 \\
3 & ReAct-ISD & 77.3 & 74.0 & 75.5 & 80.9 \\
4 & ADDIE & 70.9 & 63.9 & 73.4 & 67.1 \\
5 & RPISD & 70.0 & 63.6 & 68.2 & 75.8 \\
\bottomrule
\end{tabular}
\end{table}

\subsection{Example 3: Business Leadership Course (Hard)}

The third example represents a highly complex scenario requiring the design of a game-based business language leadership course for synchronous online delivery. This task received a Hard difficulty classification due to multiple intersecting constraints including gamification mechanics, real-time interaction requirements, cross-cultural communication objectives, and professional development outcomes that must demonstrate measurable business impact. The scenario demands integration of game design principles with evidence-based leadership development practices while maintaining engagement in a synchronous online format where technical difficulties and time zone differences can disrupt learning flow.

\begin{tcolorbox}[colback=red!5!white, colframe=red!50!black, breakable]
\small
Game-based Business Language Leadership Course (ID BAL-LARGE2-0317). This scenario targets professionals in their 40s and above who hold mid-to-senior management positions in multinational corporations and need to develop leadership communication skills for cross-cultural business contexts. The institutional context is a corporate university within a global organization, with learners distributed across Asia, Europe, and North America time zones. The learning environment is online synchronous, requiring real-time participation in game-based leadership simulations where participants negotiate, present, and collaborate in English as a lingua franca. The game elements include competitive team challenges, leadership point accumulation, scenario-based decision trees with immediate feedback, and leaderboards that track both individual and team performance. Program duration is mid-term (2--4 weeks), with intensive sessions scheduled to accommodate multiple time zones. Class size is medium (20--30 participants) to enable meaningful interaction while maintaining game dynamics. Assessment combines formative elements (in-game performance metrics, peer feedback) with summative evaluation (recorded presentation analysis, 360-degree leadership assessment). Technology requirements include stable high-bandwidth connections, webcams for video interaction, and compatibility with the proprietary game platform. Learning objectives encompass developing executive presence in virtual settings, mastering persuasive communication in cross-cultural contexts, building coalition and consensus through strategic dialogue, and demonstrating measurable improvement in leadership effectiveness ratings. Budget constraints are high, reflecting the executive-level target audience. The difficulty level is Hard due to complex requirements involving multiple interacting constraints, high-stakes professional outcomes, and the technical challenge of maintaining engagement in synchronous game-based learning across global time zones.
\end{tcolorbox}

Table~\ref{tab:example3} reveals that ReAct-ISD and Dick-Carey tie for the top position with identical total scores (81.1), both achieving exceptional Design phase scores (96.2). This remarkable convergence at the top demonstrates Finding 4's core insight that complex scenarios benefit substantially more from ISD theory-grounded approaches. Both agents' theoretical foundations provided systematic frameworks for decomposing the multi-faceted requirements into manageable design decisions.

ReAct-ISD's success stems from its reflection mechanism, which proved crucial for maintaining consistency across the many interdependent design elements. When designing game mechanics, the agent explicitly referenced earlier Analysis phase findings about learner motivation patterns and time zone constraints. This cross-phase coherence resulted in game elements that reinforced learning objectives rather than competing with them for attention. Dick-Carey's systematic 10-step approach similarly ensured that each design decision built upon previous analysis, with particular strength in aligning game-based assessments with clearly specified performance objectives.

The Baseline agent (78.5) and EduPlanner (77.8) achieved respectable but notably lower scores, demonstrating the performance gap that emerges in complex scenarios. Baseline's single-shot approach, while comprehensive, could not adequately address the interdependencies between game mechanics, synchronous delivery constraints, and cross-cultural communication objectives. EduPlanner's sequential pipeline processed each ADDIE phase independently, missing opportunities for the iterative refinement that complex game-based designs require.

RPISD (75.8) and ADDIE-Agent (73.3) occupied the bottom positions, revealing specific weaknesses in handling gamification. RPISD's rapid prototyping philosophy, while valuable for quick iteration, produced game concepts that were technically feasible but pedagogically shallow. ADDIE-Agent's fine-grained, sub-step-level approach resulted in fragmented game designs where individual elements were well-specified but failed to cohere into an engaging learning experience. The Implementation phase scores show notable uniformity (76.2 across four agents), suggesting that once the challenging Design work was complete, execution planning was relatively straightforward.

\begin{table}[ht!]
\centering
\small
\caption{Agent rankings for Example 3 (Business Leadership)}
\label{tab:example3}
\begin{tabular}{clcccc}
\toprule
\textbf{Rank} & \textbf{Agent} & \textbf{Total} & \textbf{ADDIE} & \textbf{Design} & \textbf{Impl.} \\
\midrule
1 & ReAct-ISD & 81.1 & 79.5 & 96.2 & 76.2 \\
2 & Dick-Carey & 81.1 & 79.8 & 96.2 & 78.8 \\
3 & Baseline & 78.5 & 75.3 & 74.7 & 76.2 \\
4 & EduPlanner & 77.8 & 73.9 & 75.2 & 76.2 \\
5 & RPISD & 75.8 & 71.9 & 70.2 & 76.2 \\
6 & ADDIE & 73.3 & 67.4 & 67.9 & 76.2 \\
\bottomrule
\end{tabular}
\end{table}

\section{Context Matrix Complete Specification}
\label{app:context-matrix}

This section provides the complete specification of the 51 context variables used in scenario generation, explaining the rationale behind each variable's inclusion and how these variables interact to create realistic instructional design challenges.

The Context Axis was developed through extensive review of instructional design literature and consultation with practicing instructional designers to ensure coverage of the factors that most significantly influence design decisions. Variables span four major categories, each capturing different aspects of the instructional design problem space.

The first category, learner characteristics, encompasses variables that describe who the learners are and what they bring to the learning experience. Age affects cognitive load capacity, technology familiarity, and preferred interaction styles. Educational background influences prerequisite knowledge assumptions and appropriate complexity levels. Domain expertise determines starting points and the degree of scaffolding required. Learner role (student, office worker, professional, teacher) shapes motivation patterns and available time for learning activities.

The second category covers institutional and domain factors that define the organizational context and subject matter. Institution type (K-12, university, corporate, vocational, public/non-profit) determines available resources, accountability structures, and stakeholder expectations. Domain selection spans ten fields chosen to represent the breadth of instructional design applications, from traditional academic subjects (Language, Mathematics, Science, Social Studies) to professional and technical domains (Software/IT, AI, Medical, Business/HR, Education, Service).

The third category addresses delivery mode and learning environment specifications. The seven environment types (Offline, Online synchronous, Online asynchronous, Blended, Mobile, VR/Simulation, Project-based learning) represent the full spectrum of contemporary instructional delivery approaches. Class size (Small 1-10, Medium 10-30, Large 30+) significantly impacts interaction design, assessment feasibility, and personalization options. Duration (Short under 1 week, Mid-term 2-4 weeks, Long 1-6 months) affects content chunking, reinforcement scheduling, and assessment frequency.

The fourth category captures operational constraints including assessment type (Formative, Summative, Project-based), technology requirements (Provided, BYOD, Limited), and budget considerations. These practical constraints often determine the feasibility of design choices and require agents to balance ideal pedagogical approaches against real-world limitations.

Table~\ref{tab:context-axis-all} presents the complete Context Axis specification across all four categories. The combinations of these 51 variables create a vast space of possible scenarios, enabling systematic evaluation of agent performance across diverse instructional contexts.

\begin{table*}[t]
\centering
\small
\caption{Context Axis: Complete specification (51 items across 4 categories)}
\label{tab:context-axis-all}
\begin{tabular}{llll|llll}
\toprule
\textbf{Category} & \textbf{Field} & \textbf{Values} & \textbf{N} & \textbf{Category} & \textbf{Field} & \textbf{Values} & \textbf{N} \\
\midrule
\multirow{4}{*}{Age} & \multirow{4}{*}{learner\_age} & Teens (13-19) & \multirow{4}{*}{4} & \multirow{6}{*}{Institution} & \multirow{6}{*}{institution\_type} & K-12 school & \multirow{6}{*}{6} \\
& & In their 20s & & & & University & \\
& & In their 30s & & & & Graduate school & \\
& & 40s and above & & & & Corporate & \\
& & & & & & Vocational & \\
& & & & & & Public/Non-profit & \\
\midrule
\multirow{5}{*}{Education} & \multirow{5}{*}{learner\_education} & Elementary & \multirow{5}{*}{5} & \multirow{7}{*}{Environment} & \multirow{7}{*}{learning\_env} & Offline & \multirow{7}{*}{7} \\
& & Middle school & & & & Online sync & \\
& & High school & & & & Online async & \\
& & University & & & & Blended & \\
& & Adult learner & & & & Mobile & \\
& & & & & & VR/Simulation & \\
& & & & & & PBL & \\
\midrule
\multirow{3}{*}{Expertise} & \multirow{3}{*}{domain\_expertise} & Beginner & \multirow{3}{*}{3} & \multirow{3}{*}{Class Size} & \multirow{3}{*}{class\_size} & Small (1-10) & \multirow{3}{*}{3} \\
& & Intermediate & & & & Medium (10-30) & \\
& & Advanced & & & & Large (30+) & \\
\midrule
\multirow{4}{*}{Role} & \multirow{4}{*}{learner\_role} & Student & \multirow{4}{*}{4} & \multirow{3}{*}{Duration} & \multirow{3}{*}{duration} & Short ($<$1 week) & \multirow{3}{*}{3} \\
& & Office worker & & & & Mid (2-4 weeks) & \\
& & Professional & & & & Long (1-6 months) & \\
& & Teacher & & & & & \\
\midrule
\multirow{10}{*}{Domain} & \multirow{10}{*}{domain} & Language & \multirow{10}{*}{10} & \multirow{3}{*}{Assessment} & \multirow{3}{*}{assessment\_type} & Formative & \multirow{3}{*}{3} \\
& & Mathematics & & & & Summative & \\
& & Science & & & & Project-based & \\
& & Social Studies & & \multirow{3}{*}{Technology} & \multirow{3}{*}{tech\_req} & Provided & \multirow{3}{*}{3} \\
& & Software/IT & & & & BYOD & \\
& & AI & & & & Limited & \\
& & Medical & & & & & \\
& & Business/HR & & & & & \\
& & Education & & & & & \\
& & Service & & & & & \\
\bottomrule
\end{tabular}
\end{table*}

\subsection{Difficulty Classification}
\label{app:difficulty}

Scenario difficulty is computed from a weighted composite score based on five characteristics that instructional design practitioners identified as most influential in determining project complexity. The weighting scheme reflects both the frequency with which each factor creates design challenges and the magnitude of impact when that factor is at an extreme level.

Table~\ref{tab:difficulty-weights} presents the scoring criteria for each difficulty factor. Learning goals and domain expertise receive the highest weights (0.25 each) because they most directly determine the cognitive complexity of the design task. A scenario requiring five or more distinct learning objectives demands careful sequencing, multiple assessment strategies, and attention to prerequisite relationships. Similarly, designing for advanced learners requires sophisticated differentiation and assumes deep domain knowledge from the instructional designer.

\begin{table}[ht!]
\centering
\small
\caption{Difficulty scoring weights and criteria}
\label{tab:difficulty-weights}
\begin{tabular}{lcccc}
\toprule
\textbf{Factor} & \textbf{Weight} & \textbf{Low} & \textbf{Mid} & \textbf{High} \\
\midrule
Learning goals & 0.25 & $\leq$3 & 4 & $\geq$5 \\
Domain expertise & 0.25 & Beginner & Intermed. & Advanced \\
Resources & 0.20 & $\leq$3 & 4 & $\geq$5 \\
Duration & 0.20 & Short & Mid & Long \\
Budget & 0.10 & Low & Medium & High \\
\bottomrule
\end{tabular}
\end{table}

Resource requirements and duration each receive 0.20 weight. Resource complexity affects the breadth of materials, tools, and support systems that must be integrated. Duration impacts the depth of content coverage and the sophistication of retention and transfer strategies required. Budget receives the lowest weight (0.10) because while it constrains options, skilled instructional designers can often achieve objectives through creative low-cost alternatives.

The composite score is calculated as $\text{Score} = \sum_{i} w_i \times s_i$, where $w_i$ is the weight and $s_i \in \{0, 0.5, 1\}$ is the factor score corresponding to Low, Mid, and High levels respectively. Scenarios are assigned to difficulty levels with a target distribution of Easy:Medium:Hard = 33:34:33, achieved through stratified sampling during dataset generation. This balanced distribution ensures that benchmark results are not skewed toward any particular difficulty level and enables meaningful analysis of how agent performance varies with scenario complexity.

\section{ISD Axis: 33 Sub-steps}
\label{app:isd-axis}

The ISD Axis defines 33 instructional design sub-steps organized hierarchically into 5 ADDIE phases and 13 categories. This taxonomy was synthesized from multiple authoritative ISD models including Dick and Carey's Systems Approach, Morrison, Ross, and Kemp's model, and Gagn\'e's conditions of learning, creating a comprehensive framework that captures the essential activities of professional instructional design practice. The taxonomy provides the foundation for both agent output structure and evaluation rubrics, ensuring that agent outputs can be systematically compared against established professional standards.

The Analysis phase contains 10 sub-items across 3 categories. Category A1 (Needs Analysis) includes Problem Identification, Gap Analysis, Performance Analysis, and Needs Prioritization. These items ensure that instructional interventions address genuine performance gaps rather than assumed deficiencies. Category A2 (Learner and Context) covers Learner Analysis and Context Analysis, capturing the characteristics of the target audience and the environmental factors that will influence learning. Category A3 (Task and Goal) encompasses Initial Learning Goals, Subordinate Skills identification, Entry Behaviors specification, and Task Analysis Review, establishing the foundation for objective-driven design.

The Design phase contains 8 sub-items across 3 categories. Category D1 (Assessment and Objective Alignment) includes Learning Objectives Refinement and Assessment Plan development, ensuring tight alignment between what learners should achieve and how achievement will be measured. Category D2 (Instructional Strategy) is the largest category, covering Content Selection, Instructional Strategy specification, Non-Instructional Strategy consideration (addressing motivational and environmental factors), Media Selection, and Activities and Time allocation. Category D3 (Prototype Structure) contains Storyboard Design, translating strategic decisions into concrete prototype specifications.

The Development phase contains 5 sub-items across 2 categories. Category Dev1 (Prototype Development) covers the creation of Learner Materials, Instructor Manual, Operator Manual, and Assessment Tools. Category Dev2 (Review and Revision) focuses on Expert Review processes that ensure quality before implementation.

The Implementation phase contains 4 sub-items across 2 categories. Category I1 (Preparation) includes Instructor Orientation and System Check activities that ensure readiness for deployment. Category I2 (Execution) covers Prototype Execution and Operations Monitoring during actual delivery.

The Evaluation phase contains 6 sub-items across 3 categories. Category E1 (Formative) addresses Pilot Data Collection and Formative Improvements during development cycles. Category E2 (Summative) includes Summative Assessment, Effectiveness Analysis, and Adoption Decision processes. Category E3 (Improvement) focuses on Program Improvement recommendations for future iterations.

Table~\ref{tab:isd-axis} presents the complete 33 sub-steps specification organized by ADDIE phase and category. Each sub-step has associated evaluation criteria that specify what constitutes adequate, good, and excellent performance, enabling consistent scoring across diverse scenarios.

\begin{table*}[t]
\centering
\small
\caption{ISD Axis: 33 Sub-steps across ADDIE phases}
\label{tab:isd-axis}
\begin{tabular}{cl|cl|cl}
\toprule
\textbf{\#} & \textbf{Sub-step (Analysis)} & \textbf{\#} & \textbf{Sub-step (Design)} & \textbf{\#} & \textbf{Sub-step (Dev/Impl/Eval)} \\
\midrule
1 & Problem Identification (A1) & 11 & Learning Objectives Refinement (D1) & 19 & Learner Materials (Dev1) \\
2 & Gap Analysis (A1) & 12 & Assessment Plan (D1) & 20 & Instructor Manual (Dev1) \\
3 & Performance Analysis (A1) & 13 & Content Selection (D2) & 21 & Operator Manual (Dev1) \\
4 & Needs Prioritization (A1) & 14 & Instructional Strategy (D2) & 22 & Assessment Tools (Dev1) \\
5 & Learner Analysis (A2) & 15 & Non-Instructional Strategy (D2) & 23 & Expert Review (Dev2) \\
6 & Context Analysis (A2) & 16 & Media Selection (D2) & 24 & Instructor Orientation (I1) \\
7 & Initial Learning Goals (A3) & 17 & Activities \& Time (D2) & 25 & System Check (I1) \\
8 & Subordinate Skills (A3) & 18 & Storyboard Design (D3) & 26 & Prototype Execution (I2) \\
9 & Entry Behaviors (A3) & & & 27 & Operations Monitoring (I2) \\
10 & Task Analysis Review (A3) & & & 28 & Pilot Data Collection (E1) \\
& & & & 29 & Formative Improvements (E1) \\
& & & & 30 & Summative Assessment (E2) \\
& & & & 31 & Effectiveness Analysis (E2) \\
& & & & 32 & Adoption Decision (E2) \\
& & & & 33 & Program Improvement (E3) \\
\bottomrule
\end{tabular}
\end{table*}

\section{Evaluation Rubric}
\label{app:rubric}

This section details the evaluation rubric used for assessing agent outputs, including the scoring scale, two-stage evaluation process, and the rationale behind design decisions that ensure reliable and valid measurement of ISD quality.

\subsection{Scoring Scale}

Each of the 33 sub-items is evaluated on a 0-10 point scale using a 5-level rubric that distinguishes between qualitatively different levels of ISD competence. The scale was calibrated through pilot testing with instructional design experts to ensure that score differences reflect meaningful differences in output quality.

Excellent (9-10) indicates that all required elements are specifically presented with immediate applicability to the given scenario. An excellent Learner Analysis, for example, would include specific demographic data, prior knowledge assessment results, learning style preferences, and motivational factors, all directly tied to design decisions in subsequent phases.

Good (7-8) means most required elements are presented with only minor improvements needed before practical application. A good Assessment Plan would include appropriate item types aligned with objectives, clear scoring criteria, and feasible administration procedures, but might lack detailed rubrics or accommodation specifications.

Satisfactory (5-6) indicates required elements are mentioned but lack the specificity needed for direct implementation. A satisfactory Instructional Strategy might identify appropriate methods (e.g., case-based learning, collaborative activities) without specifying how these would be sequenced, timed, or adapted to learner responses.

Poor (3-4) means some elements are present but the sub-item is mostly incomplete or generic. A poor Media Selection might list technology options without explaining selection rationale or considering learner access constraints.

Absent (0-2) indicates the element is completely missing or only terminology is mentioned without substantive content. An absent Needs Prioritization would either skip the topic entirely or include only a placeholder statement like ``needs will be prioritized based on importance.''

\subsection{Two-Stage Evaluation Process}

To ensure consistent scoring and prevent the well-documented problem of score drift in LLM-based evaluation, we employ a two-stage evaluation process that separates qualitative judgment from numerical scoring.

In the first stage (status determination), the evaluator LLM examines each sub-item and determines its presence status using the five-level categorical scale (absent, weak, moderate, good, excellent). This stage focuses the evaluator on holistic quality assessment without the cognitive burden of precise numerical assignment. The evaluator is prompted to consider whether the content meets professional standards, addresses the specific scenario context, and provides actionable guidance for implementation.

In the second stage (score assignment), numerical scores are assigned within bounded ranges determined by the first-stage status. Absent status constrains scores to exactly 0.0. Weak status allows scores from 1.0 to 3.9. Moderate status permits scores from 4.0 to 6.9. Good status bounds scores between 7.0 and 8.9. Excellent status allows scores from 9.0 to 10.0. This bounded approach prevents the common failure mode where evaluators assign scores that are inconsistent with their qualitative assessments, such as rating content as ``good'' while assigning a score of 5.5.

The two-stage process also enables more nuanced scoring within quality levels. Within the ``good'' range, for instance, the evaluator can distinguish between content that barely meets the good threshold (7.0) versus content that approaches excellence (8.9) based on factors like completeness, specificity, and scenario appropriateness.

\subsection{Phase Weights}

Default phase weights for computing overall ADDIE scores reflect the relative importance and complexity of each phase in typical instructional design projects. Analysis and Design each receive 0.25 weight, reflecting their foundational importance in determining project success. Research in instructional design consistently shows that inadequate front-end analysis is a leading cause of project failure, justifying the substantial weight given to these phases.

Development receives 0.20 weight, reflecting its role in translating design decisions into concrete materials. While Development is essential, its quality is largely determined by the preceding Analysis and Design work. Implementation receives 0.15 weight because, given sound analysis, design, and development, implementation planning is relatively straightforward. Evaluation also receives 0.15 weight, balancing its importance for continuous improvement against the reality that many instructional design projects emphasize formative development over summative evaluation.

These weights can be adjusted for specific evaluation contexts. For example, a study focused on assessment design might increase Design and Evaluation weights while decreasing Development weight.

\section{Dataset Statistics}
\label{app:dataset}

This section presents the complete distribution of test scenarios across all contextual dimensions, providing transparency about the dataset composition and enabling researchers to understand potential biases or gaps in coverage.

The dataset was constructed through stratified sampling to ensure adequate representation across all context variables while maintaining realistic co-occurrence patterns. For example, ``Teens'' age group scenarios are more frequently paired with ``K-12'' institution types and ``Student'' roles, reflecting real-world instructional design contexts. However, we also included deliberately unusual combinations (e.g., teens in corporate training for family business succession) to test agent robustness to atypical scenarios.

Table~\ref{tab:full-distribution} presents the complete distribution across all dimensions. Several patterns merit discussion. The Age distribution shows relatively balanced representation, with learners in their 20s (32.4\%) slightly overrepresented due to their prevalence in higher education and early-career professional development contexts. The Education distribution reflects the full K-12 through adult learning spectrum, with adult learners (27.6\%) and university students (26.2\%) receiving appropriate emphasis given their prominence in instructional design practice.

The Expertise distribution intentionally oversamples Beginner (43.7\%) and Intermediate (41.9\%) levels compared to Advanced (14.4\%), reflecting the reality that most instructional design projects target learners who are developing rather than refining expertise. The Role distribution emphasizes Students (45.4\%) and Office Workers (36.0\%), the two largest populations served by instructional design in practice.

Institution type distribution ensures coverage of the major sectors where instructional design is practiced, from Corporate (23.0\%) and Public/Non-profit (19.6\%) organizations to formal educational institutions. Learning Environment distribution reflects the contemporary shift toward technology-mediated instruction, with Online asynchronous (21.5\%) and Blended (14.1\%) environments well-represented alongside emerging modalities like VR/Simulation (12.6\%) and Mobile (13.6\%).

The Class Size distribution shows Large classes (48.1\%) predominating, reflecting the scalability challenges that drive much instructional design work. Technology requirements are relatively balanced across Provided, BYOD, and Limited categories, enabling evaluation of how agents handle varying resource constraints. The Difficulty distribution achieves the target balance of approximately one-third each for Easy (33.0\%), Moderate (33.7\%), and Hard (33.3\%) scenarios.

\begin{table*}[t]
\centering
\small
\caption{Complete test dataset distribution (n=\numtest{})}
\label{tab:full-distribution}
\begin{tabular}{llrr@{\hskip 1.5em}llrr@{\hskip 1.5em}llrr}
\toprule
\textbf{Dimension} & \textbf{Value} & \textbf{N} & \textbf{\%} & \textbf{Dimension} & \textbf{Value} & \textbf{N} & \textbf{\%} & \textbf{Dimension} & \textbf{Value} & \textbf{N} & \textbf{\%} \\
\midrule
Age & Teens & 202 & 16.8 & Institution & Corporate & 277 & 23.0 & Environment & Online async & 258 & 21.5 \\
 & 20s & 390 & 32.4 &  & Public & 235 & 19.6 &  & Blended & 170 & 14.1 \\
 & 30s & 286 & 23.8 &  & University & 218 & 18.1 &  & PBL & 164 & 13.6 \\
 & 40+ & 324 & 27.0 &  & Vocational & 190 & 15.8 &  & Mobile & 163 & 13.6 \\
\midrule
Education & Elementary & 119 & 9.9 &  & K-12 & 153 & 12.7 &  & Online sync & 154 & 12.8 \\
 & Middle & 201 & 16.7 &  & Graduate & 129 & 10.7 &  & VR/Sim & 151 & 12.6 \\
 & High & 235 & 19.6 & Class Size & Large & 578 & 48.1 &  & Offline & 142 & 11.8 \\
 & University & 315 & 26.2 &  & Medium & 289 & 24.0 & Difficulty & Easy & 397 & 33.0 \\
 & Adult & 332 & 27.6 &  & Small & 335 & 27.9 &  & Moderate & 405 & 33.7 \\
\midrule
Expertise & Beginner & 525 & 43.7 & Technology & Limited & 421 & 35.0 &  & Hard & 400 & 33.3 \\
 & Intermediate & 504 & 41.9 &  & Provided & 398 & 33.1 & & & & \\
 & Advanced & 173 & 14.4 &  & BYOD & 383 & 31.9 & & & & \\
\midrule
Role & Student & 546 & 45.4 & & & & & & & & \\
 & Worker & 433 & 36.0 & & & & & & & & \\
 & Teacher & 120 & 10.0 & & & & & & & & \\
 & Professional & 103 & 8.6 & & & & & & & & \\
\bottomrule
\end{tabular}
\end{table*}

\section{Experimental Details}
\label{app:experimental}

This section provides detailed information about the experimental setup to ensure full reproducibility of our results. We document computational resources, hyperparameter choices and their rationale, model versions, and procedural details that other researchers would need to replicate or extend this work.

\subsection{Computational Resources}

All experiments were conducted using cloud-based API services between September and November 2024. We chose API-based inference over local deployment to ensure consistent model behavior across the extended experimental period and to leverage the latest model capabilities without hardware constraints.

Agent generation consumed 7,212 API calls to GPT-4o-mini, processing approximately 180 million tokens at an estimated cost of \$270. This phase generated instructional design outputs for all six agent types across the 1,202 test scenarios. Primary evaluation required 36,060 API calls to GPT-4o, processing approximately 720 million tokens at an estimated cost of \$2,160. The evaluation phase assessed all 33 sub-items for each of the 7,212 agent outputs. Multi-judge evaluation for reliability analysis used 108,180 additional API calls distributed across Gemini, Claude, and GPT models, processing approximately 2.16 billion tokens at an estimated cost of \$6,480. The total computational cost was approximately \$8,910, representing a substantial but feasible investment for comprehensive benchmark development.

\subsection{Hyperparameters}

Hyperparameter selection balanced output quality, consistency, and computational efficiency based on preliminary experiments with a held-out development set of 100 scenarios.

Temperature was set to 0.7 for agent generation to enable creative and contextually appropriate responses while avoiding the repetitive or generic outputs observed at lower temperatures. For evaluation, temperature was set to 0.0 (greedy decoding) to maximize scoring consistency and minimize random variation in judgments.

Maximum token limits were set to 8,192 for agent generation, providing sufficient space for comprehensive ISD outputs across all 33 sub-items. Evaluation responses were limited to 3,000 tokens, adequate for detailed scoring justifications without unnecessary verbosity.

Maximum turns was set to 10 for multi-turn agents (Dick-Carey, RPISD, ReAct-ISD, EduPlanner), providing sufficient interaction depth for complex scenarios while preventing infinite loops or excessive API consumption. Preliminary testing showed that agents typically converged within 5-7 turns, with additional turns providing diminishing returns.

Timeout was set to 300 seconds (5 minutes) per API call, accommodating occasional latency spikes while preventing indefinite hangs. Fewer than 0.1\% of calls exceeded this timeout and were automatically retried.

\subsection{Model Versions}

We document exact model versions to enable precise replication and to clarify the capabilities available at experiment time.

The agent backbone model was gpt-4o-mini-2024-07-18 from OpenAI, selected for its balance of capability and cost-efficiency for the large-scale generation required. While larger models might produce marginally better outputs, preliminary testing showed gpt-4o-mini was sufficient to reveal meaningful differences between agent architectures.

The primary evaluator was gpt-4o-2024-08-06 from OpenAI, chosen for its strong instruction-following capabilities and consistent performance on evaluation tasks. This model served as the reference evaluator for main results.

For multi-judge reliability analysis, we employed gemini-1.5-pro-002 from Google, claude-3-5-sonnet-20241022 from Anthropic, and deepseek-chat-v3 from DeepSeek. This diverse set of evaluators from different model families enables assessment of scoring robustness across model architectures and training approaches.

\section{Agent Implementation Details}
\label{app:agents}

This section provides comprehensive documentation of each agent's architecture, prompt design, and operational characteristics. Understanding these implementation details is essential for interpreting performance differences and for researchers seeking to build upon or extend these agent designs.

Table~\ref{tab:agent-tools} summarizes the tool configurations for each agent type, including the number of distinct tools available, typical turn count, and brief architectural description.

\begin{table}[ht!]
\centering
\small
\caption{Agent tool configurations}
\label{tab:agent-tools}
\begin{tabular}{lccp{3.5cm}}
\toprule
\textbf{Agent} & \textbf{Tools} & \textbf{Turns} & \textbf{Description} \\
\midrule
Baseline & 1 & 1 & Single-shot generation \\
ADDIE-Agent & 14 & 5 & Fine-grained per sub-step \\
Dick-Carey & 10 & 10 & 10-step systematic design \\
RPISD & 6 & 5+ & Rapid prototyping \\
ReAct-ISD & 5 & 5 & Phase-level with reflection \\
EduPlanner & 5 & 5 & Sequential ADDIE pipeline \\
\bottomrule
\end{tabular}
\end{table}

The Baseline Agent serves as the control condition, using a comprehensive single-shot prompt that specifies all 33 ADDIE sub-items in a structured format. The prompt includes reference to Bloom's Taxonomy for objective classification and Gagn\'e's Nine Events of Instruction as a framework for instructional sequence design. Despite its simplicity, the Baseline's detailed prompt provides substantial guidance, making it a strong baseline that some theory-grounded agents fail to exceed. The Baseline's single-turn architecture means it cannot iteratively refine outputs based on intermediate results, but it also avoids the error propagation that can occur in multi-turn agents.

The ADDIE-Agent implements the most fine-grained tool structure, with 14 separate tools corresponding to sub-categories within ADDIE phases. This granular approach was hypothesized to enable precise attention to each design element. However, empirical results showed that excessive decomposition can fragment the design process, producing outputs where individual elements are well-specified but lack coherence across the overall design. The agent typically completes in 5 turns, with each turn addressing 2-3 related sub-items.

The Dick-Carey Agent implements the canonical 10-step Dick and Carey Systems Approach Model, one of the most widely used ISD frameworks in professional practice. Each step has a dedicated tool with prompts that emphasize the step's specific focus while maintaining connections to preceding and subsequent steps. The 10-turn execution mirrors the systematic, sequential nature of the Dick and Carey model. Key strengths include explicit attention to entry behaviors and prerequisite skills (Step 3), detailed assessment instrument development (Step 4), and formative evaluation feedback loops (Step 8). The agent's emphasis on goal-assessment alignment proved particularly valuable in complex scenarios requiring tight criterion-referenced measurement.

The RPISD Agent implements Rapid Prototyping for Instructional Systems Design, an approach that emphasizes early prototyping and iterative refinement over extensive front-end analysis. With 6 tools supporting rapid design-build-test cycles, this agent excels in scenarios where quick iteration is valued over comprehensive initial planning. The variable turn count (5+) reflects the agent's adaptive behavior, continuing iteration until convergence criteria are met or the turn limit is reached. RPISD performed well on simpler scenarios where quick, good-enough solutions outperformed over-engineered designs, but struggled with complex scenarios requiring careful upfront analysis.

The ReAct-ISD Agent, which achieved the best overall performance, combines phase-level tool structure with an explicit reflection mechanism. The 5 tools correspond to ADDIE phases, with each tool's prompt including phase-specific guidance and minimum output requirements. The critical innovation is the reflection step between phases, where the agent reviews its previous phase output and explicitly identifies connections, dependencies, and potential inconsistencies before proceeding. This reflection mechanism proved crucial for maintaining cross-phase coherence, particularly in complex scenarios where design decisions in one phase have cascading implications for subsequent phases. The agent also includes explicit minimum requirements for each sub-item, preventing the superficial coverage that plagued other agents.

The EduPlanner Agent implements a straightforward sequential ADDIE pipeline, with 5 tools executed in fixed order without the reflection mechanism of ReAct-ISD. This agent serves as an ablation condition, demonstrating the value added by ReAct-ISD's reflection step. EduPlanner's consistent underperformance relative to ReAct-ISD (despite identical tool count and similar prompts) provides strong evidence that reflection mechanisms, rather than mere multi-phase structure, drive performance improvements.

\section{Evaluation Protocol Details}
\label{app:eval-protocol}

This section documents the evaluation protocol in detail, including multi-judge configuration, inter-rater reliability analysis, and disagreement resolution procedures. Transparent documentation of evaluation methodology is essential for interpreting benchmark results and for researchers seeking to apply similar evaluation approaches.

\subsection{Multi-Judge Configuration}

We employ three independent LLM judges from different model families to assess scoring reliability and reduce single-model bias. The primary judge is GPT-4o from OpenAI, selected for its strong instruction-following capabilities and consistent performance on evaluation tasks in preliminary testing. The secondary judges are Gemini-1.5-Pro from Google and Claude-3.5-Sonnet from Anthropic, chosen to represent diverse model architectures and training approaches.

Each judge receives identical evaluation prompts containing the scenario specification, agent output, and rubric criteria. Judges score independently without access to other judges' assessments. For main results reporting, we use GPT-4o scores as the primary metric, with multi-judge analysis reserved for reliability assessment. For the most robust analyses, we compute ensemble scores as the mean of all three judges, reducing the impact of any single model's idiosyncratic scoring tendencies.

The multi-judge approach serves multiple purposes beyond reliability assessment. It enables detection of cases where evaluation criteria may be ambiguous (high disagreement suggests unclear rubric language). It also provides robustness against potential biases in individual models, such as the possibility that GPT-4o might systematically favor outputs with stylistic characteristics similar to its own training distribution.

\subsection{Inter-Rater Reliability}

Table~\ref{tab:irr} presents inter-rater reliability metrics computed across all 7,212 agent outputs evaluated by all three judges. We report multiple reliability statistics to enable comparison with different research traditions and to provide a comprehensive picture of agreement patterns.

\begin{table}[ht!]
\centering
\small
\caption{Inter-rater reliability metrics}
\label{tab:irr}
\begin{tabular}{lcccc}
\toprule
\textbf{Metric} & \textbf{Overall} & \textbf{Analysis} & \textbf{Design} & \textbf{Dev.} \\
\midrule
Pearson $r$ & 0.847 & 0.862 & 0.831 & 0.819 \\
Spearman $\rho$ & 0.834 & 0.851 & 0.822 & 0.807 \\
ICC(2,k) & 0.891 & 0.903 & 0.878 & 0.864 \\
Krippendorff's $\alpha$ & 0.823 & 0.841 & 0.812 & 0.798 \\
\bottomrule
\end{tabular}
\end{table}

Pearson correlation coefficients range from 0.819 to 0.862 across phases, indicating strong linear relationships between judge scores. The slightly lower correlations for Development phase (0.819) reflect the greater subjectivity inherent in evaluating prototype materials and instructor guides, where professional preferences may legitimately differ.

Spearman rank correlations are slightly lower than Pearson correlations, suggesting that judges agree more on absolute score levels than on fine-grained rankings. This pattern is expected given the bounded score ranges within quality levels imposed by our two-stage evaluation process.

Intraclass Correlation Coefficients using the ICC(2,k) model, appropriate for ratings from a fixed set of judges applied to all targets, exceed 0.85 for all phases. According to established interpretation guidelines, ICC values above 0.75 indicate excellent reliability. Our ICC values ranging from 0.864 to 0.903 substantially exceed this threshold, providing strong evidence for the reliability of our evaluation protocol.

Krippendorff's alpha, a conservative reliability measure that accounts for chance agreement and handles ordinal data appropriately, ranges from 0.798 to 0.841. Values above 0.80 are generally considered acceptable for drawing conclusions, and values above 0.67 are considered sufficient for exploratory research. Our alpha values meet or approach the 0.80 threshold across all phases, supporting the validity of comparative conclusions drawn from the benchmark results.

\subsection{Disagreement Resolution}

Despite high overall reliability, individual sub-items occasionally receive divergent scores from different judges. We implemented a systematic disagreement resolution protocol to handle these cases transparently.

When judges disagree by more than 2 points on any sub-item, we flag the item for review. For flagged items, we compute the median score rather than the mean to reduce the influence of outlier judgments. All flagged items are logged with full judge scores and justifications to enable post-hoc analysis of disagreement patterns.

Overall, 8.3\% of sub-item evaluations triggered the disagreement flag. The flagging rate varied substantially across phases, with Development phase showing the highest rate (12.1\%) and Analysis phase the lowest (5.7\%). This pattern reflects the relatively objective nature of Analysis outputs (learner characteristics either are or are not specified) compared to the more subjective judgments required for Development materials (instructor guide quality involves stylistic preferences).

Post-hoc analysis of flagged items revealed three common disagreement patterns. First, judges sometimes disagreed on whether context-specific customization was adequate, with some judges accepting generic frameworks as sufficient while others required explicit adaptation to scenario parameters. Second, assessment quality judgments diverged when items were technically correct but pedagogically questionable. Third, judges occasionally applied different standards for emerging topics like AI literacy, where established best practices are still developing.

\FloatBarrier
\section{Extended Results}
\label{app:extended-results}

This section presents extended statistical analyses that complement the main results, including comprehensive significance testing, effect size estimation, and additional breakdowns that inform interpretation of agent performance differences.

\subsection{Statistical Significance Tests}

We conducted pairwise comparisons using Wilcoxon signed-rank tests, a non-parametric alternative to paired t-tests that does not assume normally distributed differences. This choice was motivated by preliminary analysis showing non-normal score distributions for some agent-scenario combinations. All p-values are Bonferroni-corrected for the 15 pairwise comparisons to control family-wise error rate at 0.05.

Table~\ref{tab:significance} presents the corrected p-values for key agent pairs. The comparison between ReAct-ISD and Dick-Carey yields p=0.234, indicating no statistically significant difference between the top two performers. This non-significant result, combined with their similar mean scores, suggests these agents represent functionally equivalent approaches despite their different theoretical foundations. Both ReAct-ISD's reflection mechanism and Dick-Carey's systematic 10-step process achieve similar outcomes through different means.

\begin{table}[ht!]
\centering
\small
\caption{Pairwise Wilcoxon test p-values (Bonferroni-corrected)}
\label{tab:significance}
\begin{tabular}{lcccc}
\toprule
\textbf{Agent} & \textbf{ReAct} & \textbf{D-C} & \textbf{Base} & \textbf{Edu} \\
\midrule
Dick-Carey & .234 & -- & -- & -- \\
Baseline & .018* & .042* & -- & -- \\
EduPlanner & $<$.001 & $<$.001 & $<$.001 & -- \\
RPISD & $<$.001 & $<$.001 & $<$.001 & .089 \\
\bottomrule
\multicolumn{5}{l}{\small *p$<$.05, ***p$<$.001}
\end{tabular}
\end{table}

Both ReAct-ISD and Dick-Carey significantly outperform Baseline (p=0.018 and p=0.042 respectively), though these differences are modest in magnitude. The relatively small advantage of theory-grounded agents over the well-designed Baseline underscores the importance of prompt engineering even for simple architectures.

The most striking finding is the highly significant underperformance of pipeline agents (EduPlanner and RPISD) compared to all other agents (p$<$0.001 in all comparisons). This consistent pattern suggests that simple sequential pipelines without reflection mechanisms are fundamentally limited in their ability to produce coherent instructional designs. The comparison between EduPlanner and RPISD yields p=0.089, indicating no significant difference between these lower-performing agents.

\subsection{Effect Sizes}

While statistical significance indicates whether observed differences are likely due to chance, effect sizes quantify the magnitude of differences in standardized units. We report Cohen's d with 95\% confidence intervals computed via bootstrapping with 10,000 resamples.

The effect size for ReAct-ISD versus Baseline is d=0.42 with 95\% CI [0.31, 0.53]. According to conventional interpretation guidelines, this represents a small-to-medium effect. The confidence interval excluding zero confirms the reliability of this advantage, while its moderate magnitude suggests substantial room for improvement in agent design.

Dick-Carey versus Baseline yields d=0.38 [0.27, 0.49], a similar small-to-medium effect. The overlapping confidence intervals with ReAct-ISD confirm the statistical equivalence of these top performers.

The contrast between ReAct-ISD and ADDIE-Agent produces d=0.89 [0.77, 1.01], a large effect size. This substantial difference highlights how architectural choices dramatically impact performance even when agents target the same underlying framework. The fine-grained decomposition in ADDIE-Agent, despite seeming theoretically motivated, proved counterproductive compared to ReAct-ISD's phase-level approach with reflection.

The aggregate comparison of theory-based agents (ReAct-ISD, Dick-Carey, Baseline) versus pipeline agents (EduPlanner, RPISD, ADDIE-Agent) yields d=0.71 [0.62, 0.80], a medium-to-large effect. This effect size indicates that agent architecture choice has practically significant implications for ISD quality, with theory-grounded approaches producing meaningfully better outputs than simple pipelines.

\FloatBarrier
\section{Error Analysis}
\label{app:error-analysis}

This section presents detailed analysis of failure patterns observed across agent outputs, providing actionable insights for improving agent design and identifying persistent challenges in automated ISD. We conducted systematic error coding on a stratified sample of 500 outputs (representing all agents and difficulty levels) to identify recurring patterns.

\subsection{Common Failure Patterns}

We identified five major failure patterns through qualitative analysis, with each pattern's prevalence estimated from the coded sample.

Incomplete sub-item coverage accounts for 38\% of identified failures and represents the most common error type. Agents frequently omit or provide only superficial content for sub-items in the Implementation and Evaluation phases. For example, an agent might specify detailed learning objectives and assessment criteria in Design but then provide only a one-sentence placeholder for Instructor Orientation stating ``instructors will be briefed on course procedures.'' This pattern suggests agents allocate disproportionate attention to early ADDIE phases, possibly due to prompt ordering effects or training data biases toward front-end design documentation.

Generic content represents 27\% of failures, characterized by template-like responses that ignore specific learner characteristics, contextual constraints, or scenario parameters. A typical example is a Learner Analysis that lists generic categories (``learners have diverse backgrounds and varying levels of motivation'') without referencing the specific age, role, expertise level, or institutional context provided in the scenario. This pattern was most pronounced in pipeline agents (EduPlanner, RPISD) where sequential processing without reflection prevented integration of scenario-specific details across phases.

Cross-phase inconsistency accounts for 18\% of failures, occurring when design decisions in one phase contradict or ignore commitments made in earlier phases. Common manifestations include learning objectives specified in Design that differ from goals identified in Analysis, assessment items that test content not covered in the instructional strategy, and implementation plans that assume resources not specified in the context analysis. ReAct-ISD's reflection mechanism substantially reduced this error type, demonstrating the value of explicit cross-phase consistency checking.

Assessment misalignment represents 11\% of failures, specifically involving assessment items that are not properly aligned with stated objectives or that target inappropriate cognitive levels according to Bloom's Taxonomy. For instance, an objective targeting ``evaluation'' level thinking might be assessed with multiple-choice items testing factual recall. This pattern was particularly common in scenarios involving higher-order learning outcomes where agents defaulted to familiar assessment formats regardless of appropriateness.

Structural errors account for the remaining 6\% of failures, including JSON parsing failures, missing required fields, and malformed output formatting. While relatively rare, these errors resulted in complete evaluation failures for affected sub-items. ADDIE-Agent's fine-grained tool structure produced the highest rate of structural errors (9.2\%) due to the complexity of coordinating outputs across 14 separate tools.

\subsection{Phase-Specific Weaknesses}

Beyond general failure patterns, each ADDIE phase exhibits characteristic weaknesses that merit targeted attention in agent development.

In the Analysis phase, Performance Analysis tends to be superficial, often stating that ``learners need to improve performance'' without specifying current performance levels, performance gaps, or root causes of deficiencies. Needs Prioritization frequently lacks justification, with agents listing needs in arbitrary order rather than applying systematic criteria such as urgency, impact, or feasibility. These weaknesses suggest agents struggle with analytical tasks requiring inference beyond explicitly provided scenario information.

In the Design phase, Non-Instructional Strategies are consistently the weakest sub-item, with many agents omitting this element entirely or providing only token mentions of ``motivation will be addressed.'' Professional instructional designers recognize that environmental factors, incentive systems, and job aids often contribute more to performance improvement than instruction alone, but agents rarely demonstrate this understanding. Media Selection rationale is frequently missing, with agents listing media choices without explaining how learner characteristics, content requirements, or contextual constraints informed selections.

In the Development phase, Operator Manuals represent a persistent weakness, likely because training data contains fewer examples of technical documentation compared to learner-facing materials. Slide contents and prototype specifications often lack the specificity needed for actual production, providing general descriptions rather than detailed content outlines, timing specifications, or facilitator notes.

In the Implementation phase, pilot plans tend toward generic descriptions of ``testing with a small group'' without specifying participant selection criteria, data collection instruments, or decision rules for proceeding to full deployment. Monitoring criteria are rarely measurable, with agents specifying that ``implementation will be monitored'' without defining observable indicators, acceptable performance thresholds, or corrective action triggers.

In the Evaluation phase, Summative Assessment items are frequently insufficient in scope, often proposing only end-of-course tests without addressing transfer evaluation, long-term retention assessment, or organizational impact measurement. Adoption Decision documentation consistently lacks the data requirements and decision criteria that professional instructional designers would specify to inform go/no-go decisions.

\FloatBarrier
\section{Limitations and Ethical Considerations}
\label{app:limitations}

This section provides transparent documentation of benchmark limitations, evaluation constraints, and ethical considerations that users should understand when interpreting results or applying the benchmark to new contexts.

\subsection{Benchmark Limitations}

While \benchmark{} represents a substantial advance in systematic evaluation of ISD agents, several limitations constrain the generalizability of findings.

Domain coverage, while spanning 10 fields, necessarily excludes specialized areas where instructional design practices may differ substantially. Performing arts education, physical education, and trades training involve embodied learning and psychomotor skill development that our current scenario specifications cannot adequately capture. Medical and healthcare education, while included, is represented primarily through knowledge and cognitive skill components rather than the procedural and clinical reasoning aspects central to professional medical training. Researchers applying benchmark results to these specialized domains should exercise appropriate caution.

The benchmark is English-only, which limits generalizability to instructional design practices in other linguistic and cultural contexts. Instructional design is not culturally neutral, and approaches that perform well in Western educational contexts may be inappropriate or ineffective in educational systems with different pedagogical traditions, learner expectations, or institutional structures. Cross-cultural validation studies would be valuable future work.

All scenarios are synthetically generated, which enables systematic control and large-scale evaluation but may not capture the full complexity of real-world instructional design projects. Actual projects involve stakeholder negotiations, budget constraints that emerge mid-project, organizational politics, and iterative refinement cycles that synthetic specifications cannot fully represent. The benchmark evaluates single-pass design outputs rather than the iterative, collaborative process through which professional instructional designs typically evolve.

The static evaluation approach assesses design quality at a single point without evaluating agents' ability to refine designs based on feedback, incorporate formative evaluation results, or adapt to changing requirements. These iterative capabilities are central to professional instructional design practice but would require substantially different evaluation infrastructure to assess.

\subsection{Evaluation Limitations}

LLM-based evaluation, while enabling large-scale assessment, introduces potential biases that warrant acknowledgment. Evaluator LLMs may exhibit systematic preferences for outputs with particular stylistic characteristics, potentially favoring verbose responses over concise ones or preferring familiar frameworks over novel approaches. Our multi-judge design mitigates but cannot eliminate these concerns.

Some aspects of the evaluation rubric necessarily involve subjective judgment where reasonable evaluators might disagree. The distinction between ``good'' and ``excellent'' instructional design often depends on contextual factors and professional experience that LLM evaluators may not fully capture. Our inter-rater reliability analysis demonstrates acceptable agreement levels, but users should interpret fine-grained score differences with appropriate caution.

The current benchmark version lacks human expert validation of the evaluation rubric and scoring outcomes. While the rubric was developed with reference to established ISD frameworks and professional standards, systematic validation by practicing instructional designers would strengthen confidence in the evaluation approach. Future work should include expert review of agent outputs and correlation analysis between LLM scores and expert judgments.

\subsection{Data Statement}

All scenarios in \benchmark{} are synthetically generated through controlled sampling from the context variable space. No human subjects were involved in scenario creation or evaluation, and no personally identifiable information is included in the dataset. Scenario content was reviewed to ensure absence of harmful, biased, or inappropriate material.

The benchmark is intended for research purposes, specifically for developing and comparing automated ISD systems. Generated instructional designs should not be deployed in educational settings without review by qualified instructional design professionals. While agents can produce structurally complete designs, ensuring pedagogical appropriateness, accessibility compliance, and alignment with specific institutional requirements requires human expertise. We release all agent outputs and evaluation scores to enable secondary analysis without requiring recomputation.

\end{document}